\documentstyle[aps,prb,floats,epsfig]{revtex}

\begin{document}

\draft

\title{$1/N$ expansion for two-dimensional quantum ferromagnets}
\author{Carsten Timm, S.M. Girvin, and Patrik Henelius}
\address{Department of Physics, Indiana University, Bloomington,
Indiana 47405}
\date{January 20, 1998}

\maketitle

\begin{abstract}
The magnetization of a two-dimensional ferromagnetic Heisenberg model,
which represents a quantum Hall system at filling factor $\nu=1$,
is calculated employing a large $N$ Schwinger boson approach.
Corrections of order $1/N$ to the mean field ($N=\infty$) results for
both the SU($N$) and the O($N$) generalization of the bosonized
model are presented. The calculations are discussed in detail and
the results are compared with quantum Monte Carlo simulations
as well as with recent experiments. The SU($N$) model describes both
Monte Carlo and experimental data well at low temperatures,
whereas the O($N$) model is much better at moderate and high
temperatures.
\end{abstract}


\widetext

\section{Introduction}
\label{intro}

Progress in materials synthesis has allowed to study a variety of
two-dimensional (2D) systems such as thin films, surfaces, and
semiconductor quantum wells. These systems as well as the nearly
2D cuprates have lead to much interest in
2D quantum magnetism. It has been found that
2D electron gases in quantum wells in the quantum Hall
regime are novel itinerant ferromagnets.\cite{Sondhi,Cote,QHferro}
The strong external magnetic field quenches the kinetic
energy, leading to widely separated Landau levels, but 
because of band structure effects it couples only
weakly to the electron spins. Thus low-energy spin fluctuations
play an important role.

These 2D continuum ferromagnets exhibit topological excitations called
skyrmions\cite{Sondhi,Moon} in analogy to the Skyrme model of nuclear
physics.\cite{Skyrme,Raja} In the quantum Hall system these
excitations carry electrical charge.\cite{Sondhi,Moon} At filling
factor $\nu=1$, {\it i.e.}, if the spin-up states in the lowest Landau
level are just filled, skyrmions only appear as thermal excitations of
the form of skyrmion-antiskyrmion pairs. At filling factors away from
unity, however, skyrmions appear even in the ground
state.\cite{QHground} At all filling factors, low-energy spin
fluctuations are also present. The combination of spin fluctuations
and skyrmions dramatically alters the
magnetization\cite{Barrett,Cote,Manfra} and the specific
heat.\cite{Bayot}

For a quantitative understanding it is useful to first
study the case of $\nu=1$ to isolate the effect of low-energy
spin fluctuations, which are expected to be well described by a
Heisenberg model, at least at low enough temperatures. At
higher temperatures higher order gradient terms neglected in
the Heisenberg model could become important.
Renormalization group arguments\cite{RS} show that in $D=2-\epsilon$
dimensions the magnetization $M$ of the quantum Hall ferromagnet at
$\nu=1$ is a universal function, $M/S = f(JS^2/T,B/T)$, where
$S$ is the spin, $J$ is the exchange coupling, and $B$ is the external
magnetic field. For $D=2$ this universality is violated by
logarithmic corrections.\cite{RS} In the Heisenberg model
the magnetization only depends on the {\it three\/} dimensionless
quantities $S$, $J/T$, and $B/T$.
Read and Sachdev\cite{RS} have evaluated the
magnetization using SU($N$) and O($N$) Schwinger boson
formulations of mean field (MF) theory, {\it i.e.}, $N\to\infty$,
for the Heisenberg model.
In a recent communication\cite{letter} we have presented analytic
results for the leading $1/N$ corrections to the magnetization and
results of extensive quantum Monte Carlo simulations.
In the present paper we present details of the $1/N$ theory. Details
of the Monte Carlo simulations are given elsewhere.\cite{PH}
An alternative microscopic approach which includes spin-wave
corrections to the electronic self-energy
has also recently been developed.\cite{KM}

Schwinger boson theories\cite{Schwinger,SbSoS} have
proved useful in finding MF theories
which respect the symmetry of the Hamiltonian.
Formal results to any order in $1/N$ have also been
obtained.\cite{Auerbach,Starykh} However, numerically evaluating the
first order ($1/N$) corrections is not an easy task.
Trumper {\it et al}.\cite{TMGC} have evaluated various ground-state
quantities of a frustrated antiferromagnet
in the absence of external fields.
Although they are not using the large $N$ formalism, their
method is equivalent to a $1/N$ expansion to first order.

There are a number of subtle pitfalls in the $1/N$ calculations,
{\it e.g.}, regarding normal ordering of operators. It seems
worthwhile to present the calculations in some detail for the benefit
of readers interested in using $1/N$ expansion methods. We also hope
to make the physical interpretation of these theories clearer and shed
some light on the level of accuracy of $1/N$ expansions.

In the following we give an overview of this work. First,
the Heisenberg Hamiltonian is mapped onto an equivalent
boson system. There are several ways of doing this.
One is the Holstein-Primakoff representation,\cite{HP}
which has a number of disadvantages, {\it e.g.}, the square root
of operators it introduces, and we do
not employ it here. Instead we introduce Schwinger
bosons\cite{Schwinger} in two different ways. The first, presented in
Sec.~\ref{2.1}, makes use of the SU(2) symmetry
in spin space of the Heisenberg model (which is
explicitly broken by an external field). The second utilizes the
local equivalence between the groups SU(2) and O(3) to write
down an equivalent O(3) boson model (Sec.~\ref{3.1}). Subsequently, the
two models are generalized to SU($N$) and O($N$), respectively,
which contain $N$ bosons at each site. At this point a remark
may be in order on what we do {\it not\/} mean by the O($N$) model.
It is not an $N$ component vector model, {\it e.g.}, an $N$ component
quantum non-linear sigma model. Rather, the spin operators are
generators of the Lie group O($N$). Only for $N=3$ are the
generators antisymmetric $3\times3$ matrices, which are
dual to (axial) vectors. Thus our results are
not easily compared to expansions in the number
of components of the spin vectors, as developed by
Garanin\cite{Garanin} for a classical system.

It is now possible to expand in $1/N$ as a small
parameter. MF theory becomes exact for
both SU($\infty$) and O($\infty$).\cite{Auerbach}
The $1/N$ expansion is a saddle-point expansion around this MF
solution, not a perturbative expansion in the interaction.
For this reason it is, in principle, equally valid at
all temperatures. Also, it does respect the symmetry of
the Heisenberg model. This property makes
even the MF results qualitatively correct. In particular,
the absence of long range order if no external field is present
is correctly predicted.
After rederiving the MF magnetization in Sec.~\ref{2.2} for the
SU($N$) model and in Sec.~\ref{3.2} for O($N$),
we calculate the
$1/N$ corrections using a diagrammatic approach\cite{Auerbach}
(Secs.~\ref{2.3} and \ref{3.3}).
These corrections take fluctuations around the MF result into account.
We will make use of gauge invariance to simplify our task.
Here we also have to discuss the effect of normal ordering.
In principle, terms to any order in $1/N$ can be obtained
in the same way.

The system without exchange interaction can be solved exactly for any
value of the spin $S$ and for any $N$ in both the SU($N$) and the
O($N$) model. It
can be used to check the $1/N$ expansion. However, the interaction
introduces a number of additional complications.

\section{SU($N$) model}

\subsection{General considerations}
\label{2.1}

We start from a Heisenberg model with nearest-neighbor
interaction on a square lattice in a constant magnetic field,
\begin{equation}
H = -J \sum_{\langle ij\rangle} {\bf S}(i)\cdot{\bf S}(j)
  - B \sum_i S^z(i) ,
\label{11H1}
\end{equation}
where the sum over $\langle ij\rangle$ is over all nearest-neighbor
bonds. A factor of $g\mu_B$ has been absorbed into the field $B$.
The total spin at each
site is $S$; ${\bf S}(i)\cdot{\bf S}(i) = S(S+1)$. We
express the spins in terms of Bose operators using a Schwinger
boson representation, where two Bose fields $a$ and $b$ are introduced
according to\cite{Schwinger,AA}
\begin{equation}
S^+ = a^\dagger b , \quad
S^- = b^\dagger a , \quad
S^z = (a^\dagger a-b^\dagger b)/2 .
\label{11d1}
\end{equation}
To restrict the Hilbert space to the physical states,
the constraint $a^\dagger a + b^\dagger b = 2S$
is introduced, which corresponds to ${\bf S}\cdot{\bf S} = S(S+1)$ for
the original Hamiltonian. The boson Hamiltonian is
\begin{eqnarray}
H & = & -\frac{J}2 \sum_{\langle ij\rangle} \Big[
    a^\dagger(i)a(i)a^\dagger(j)a(j) + a^\dagger(i)b(i)b^\dagger(j)a(j)
  \nonumber \\
& & {}+ b^\dagger(i)a(i)a^\dagger(j)b(j)
  + b^\dagger(i)b(i)b^\dagger(j)b(j) \Big]  \nonumber \\
& & {}- \frac{B}2 \sum_i \Big[a^\dagger(i)a(i)
  - b^\dagger(i)b(i)\Big] ,
\end{eqnarray}
neglecting a constant. For this Hamiltonian to be equivalent to the
Heisenberg model, the spin operators expressed in terms of bosons
have to have the correct commutation relations.
This is easily shown to be the case.

Utilizing the SU(2) symmetry group of the spins
we write the Hamiltonian in a more compact
form by first defining a SU(2) spin matrix
\begin{equation}
\text{\sf S} \equiv \left( \begin{array}{cc}
                      a^\dagger a & a^\dagger b \\
		      b^\dagger a & b^\dagger b
                \end{array} \right) ,
\label{11S3}
\end{equation}
with the constraint $\text{Tr}\,\text{\sf S}=2S$. The
Hamiltonian is
\begin{equation}
H = -\frac{J}2 \sum_{\langle ij\rangle} S_\beta^\alpha(i)
    S_\alpha^\beta(j)
  - \frac{B}2 \sum_i (\sigma^z)_\beta^\alpha S_\alpha^\beta(i) ,
\label{11H3}
\end{equation}
where $\sigma^z$ is a Pauli matrix and summation over repeated indices
is implied. Here, the spin matrices $\text{\sf S}(i)$ should be
infinitesimal generators of the SU(2) group, {\it i.e.}, elements of
the corresponding algebra. This is not the case since the generators
are traceless. However, if we had defined $\text{\sf S}$ as an element
of the algebra, the Hamiltonian would only change by a constant and we
use the more convenient definition~(\ref{11S3}).

The group SU(2) is generalized to SU($N$) for any
even $N$. The generalization of the Hamiltonian (\ref{11H3}) is
\begin{equation}
H = -\frac{J}{N} \sum_{\langle ij\rangle} S_\beta^\alpha(i)
    S_\alpha^\beta(j)
  - \frac{B}2 \sum_i h_\beta^\alpha S_\alpha^\beta(i) ,
\label{11H6}
\end{equation}
where $\text{\sf S}$ and $h$ are $N\times N$ hermitian matrices and
$\text{\sf S}$ is subject to the constraint
$\text{Tr}\,\text{\sf S} = NS$. We choose
$h_\beta^\alpha = \delta_{\alpha\beta} (-1)^{\alpha+1}$
so that we regain the SU(2) model for $N=2$. The Schwinger boson
representation now requires $N$ boson species $b_\alpha$,\cite{AA}
$S_\beta^\alpha = b_\alpha^\dagger b_\beta$, and the constraint is
\begin{equation}
b_\alpha^\dagger b_\alpha = NS .
\label{11c6}
\end{equation}
We now go over to the continuum for mathematical convenience.
The continuum model may actually give a better description of
itinerant magnets but is harder to compare to Monte Carlo simulations
on a lattice. Up to a constant we obtain
\begin{equation}
H = \int d^2r\, \left[ \frac{J}{2N} (\partial_j S_\beta^\alpha)
  (\partial_j S_\alpha^\beta) - \frac{B}{2a^2} h_\beta^\alpha
    S_\alpha^\beta \right] ,
\label{11H7}
\end{equation}
where $b_\alpha({\bf r})$ is a continuous Bose field with the
commutator
$[b_\alpha({\bf r}),b_\beta^\dagger({\bf r}')]
  = a^2\delta_{\alpha\beta}\delta({\bf r-r}')$,
$\partial_j$ is the two-component gradient, $a$ is the lattice
constant, and summation over $j$ is implied. After bosonization we find
\begin{eqnarray}
H & = & \int d^2r\, \bigg[ JS (\partial_j b_\alpha^\dagger)
    (\partial_j b_\alpha)
  - \frac{J}{N} b_\alpha^\dagger(\partial_j b_\beta^\dagger)
    b_\beta(\partial_j b_\alpha) \nonumber \\
& & {}- \frac{B}{2a^2} h_\beta^\alpha b_\beta^\dagger b_\alpha \bigg] ,
\end{eqnarray}
which is normal ordered, as necessary for the functional integral.
This is basically the Hamiltonian of the complex projective $CP^{N-1}$
model.\cite{Starykh} We have used the fact that the lattice
Hamiltonian (\ref{11H6}) can be normal ordered trivially since spins
at different sites commute so that
$S_\beta^\alpha(i) S_\alpha^\beta(j)
  =\;:S_\beta^\alpha(i) S_\alpha^\beta(j):$~,
where $:\;:$ denotes normal ordering.

Now we write down the partition function as a coherent state functional
integral, where the Bose fields are replaced by complex fields,
\begin{equation}
Z = \int D^2b_\alpha\,D\lambda \exp\!\left( -\frac1{\hbar}
  \int_0^{\hbar\beta}
  \!\!\!d\tau \int d^2r\, {\cal L}[b;\lambda] \right) ,
\label{11Z1}
\end{equation}
where the functional integral takes each $b_\alpha({\bf r},\tau)$ over
the whole complex plane and each $\lambda({\bf r},\tau)$ parallel to
the imaginary axis (a constant real part is irrelevant).
Here and in the following we neglect constant factors in $Z$.
$\tau$ is the imaginary time, $\beta$ is the inverse temperature,
and ${\cal L}$ is the Lagrangian
\begin{eqnarray}
{\cal L} & = & \frac{\hbar}{a^2} b_\alpha^\ast \partial_0 b_\alpha
  + JS (\partial_j b_\alpha^\ast)(\partial_j b_\alpha)
  - \frac{J}{N} b_\alpha^\ast(\partial_j b_\beta^\ast)
    b_\beta(\partial_j b_\alpha) \nonumber \\
& & {}- \frac{B}{2a^2} h_\beta^\alpha b_\beta^\ast b_\alpha
  + \lambda b_\alpha^\ast b_\alpha - NS \lambda .
\label{11L2}
\end{eqnarray}
The first term is the usual Berry phase ($\partial_0$ is the time
derivative) and the last two terms come from the constraint using the
identity $2\pi\delta(\phi) = \int_{-\infty}^\infty dx\,e^{ix\phi}$.
$\lambda$ is a Lagrange multiplier at each point $({\bf r},\tau)$.

To decouple the quartic term we introduce a
Hubbard-Stratonovich field ${\bf Q}({\bf r},\tau)$: Since
\begin{eqnarray}
& & \int DQ_j \exp\!\Bigg( -\frac1{\hbar} \int_0^{\hbar\beta}\!\!\!d\tau
  \int d^2r
  \frac{J}{N} \big[-iNQ_j - (\partial_j b_\alpha^\ast)b_\alpha \big]
  \nonumber \\
& & \qquad\times \big[iNQ_j - b_\beta^\ast(\partial_j b_\beta)\big] \Bigg)
\end{eqnarray}
is independent of $b_\alpha$, we can multiply the partition function
with this expression. $Q_j$ can be chosen
real since an imaginary part of $Q_j$ would not couple to the
$b_\alpha$ fields
because $b_\beta^\ast\partial_j b_\beta$ is purely imaginary. We get
\begin{equation}
Z = \int D^2b_\alpha\,D\lambda\,DQ_j \exp\!\left( -\frac1{\hbar}
  \int_0^{\hbar\beta}\!\!\!d\tau \int d^2r\,
    {\cal L}'[b;\lambda,{\bf Q}] \right) 
\label{11Z4}
\end{equation}
with
\begin{eqnarray}
{\cal L}' & = & \frac{\hbar}{a^2} b_\alpha^\ast \partial_0 b_\alpha
  + JS (\partial_j b_\alpha^\ast)(\partial_j b_\alpha)
  + NJ Q_jQ_j  \nonumber \\
& & {}+ iJ Q_j b_\alpha^\ast(\partial_j b_\alpha)
    - iJ Q_j (\partial b_\alpha^\ast)b_\alpha  \nonumber \\
& & {}- \frac{B}{2a^2} h_\beta^\alpha b_\beta^\ast b_\alpha
  + \lambda b_\alpha^\ast b_\alpha - NS \lambda .
\label{11L4}
\end{eqnarray}
We see that ${\bf Q}$ is a {\it gauge field\/}:
If we multiply all $b_\alpha$ by a local phase factor,
$b_\alpha({\bf r},\tau) \to e^{i\theta({\bf r},\tau)}
  b_\alpha({\bf r},\tau)$,
we reobtain the Lagrangian (\ref{11L4}) by letting
$Q_j \to Q_j + S\,\partial_j\theta$.
We know from gauge theory that ${\bf Q}$ contains more information
than is physically relevant; we have the freedom to
choose a gauge. We use a transverse gauge,
\begin{equation}
\partial_jQ_j = 0 .
\label{11g2}
\end{equation}
Of course we obtain the same results if we do not fix the gauge.
The gauge freedom then leads to the appearance of zero modes,
which turn out not to enter in the magnetization.

\subsection{Mean field theory}
\label{2.2}

Up to this point the treatment has been exact.
In the following we derive mean field (MF) results, which are exact
for $N\to\infty$ and approximate for finite $N$. This approximation
is not the same as standard MF theory for the Heisenberg model. As we
will see, SU($N$) MF theory captures the low-energy spin-wave physics
of the Heisenberg model and correctly predicts the absence of long
range order at finite temperatures.

The MF approximation is the leading order of a stationary phase
approximation for the SU($N$) partition function.
The MF solution is assumed to be homogeneous and static, {\it i.e.},
${\bf Q}$ and $\lambda$ are assumed to be constant.
This assumption is not justified for all systems,\cite{A:const}
it should hold in ferromagnets, though.\cite{Auerbach}
The MF values $\overline{\bf Q}$
and ${\overline{\lambda}}$ are chosen in such a way that the
MF free energy $F_0$ has a saddle point.
If we set $\lambda$ to its MF value the constraint
(\ref{11c6}) is no longer satisfied locally but only on average.
In order to diagonalize the action
we introduce Fourier transforms of the $b_\alpha$ fields,
\begin{equation}
b_\alpha({\bf r},\tau) = \frac{a^2}{2\pi} \int d^2k \sum_{i\omega_n}
  e^{i{\bf k}\cdot{\bf r}-i\omega_n\tau} b_\alpha({\bf k},i\omega_n) ,
\label{11fou1}
\end{equation}
where $i\omega_n = i2\pi n/\hbar\beta$ are bosonic Matsubara
frequencies. From now on summation over indices is written out.
With Eq.~(\ref{11L4}) and the definition
$h_\beta^\alpha = \delta_{\alpha\beta} (-1)^{\alpha+1}$
the MF partition function is
\begin{eqnarray}
Z_0 & = & \int D^2b_\alpha({\bf k},i\omega_n)\; \exp\!\Bigg( -{\cal N}
  N\beta J \overline{\bf Q}\cdot\overline{\bf Q} a^2 \nonumber \\
& & {}+ {\cal N} a^2NS \beta{\overline{\lambda}}
  - \int d^2k \sum_{i\omega_n} {\cal L}''_0[b] \Bigg) ,
\label{12Z3}
\end{eqnarray}
where ${\cal N}$ is the total number of sites and
\begin{eqnarray}
{\cal L}''_0 & = & \beta a^2 \sum_\alpha \bigg(
  - i\hbar\omega_n + JS k^2a^2 - 2J \overline{\bf Q}\cdot{\bf k} a^2
  \nonumber \\
& & {}- \frac{B}2 h_\alpha^\alpha + a^2 {\overline{\lambda}} \bigg)
  \; b_\alpha^\ast({\bf k},i\omega_n)b_\alpha({\bf k},i\omega_n) .
\label{12L3}
\end{eqnarray}
We introduce a number of new symbols,
\begin{equation}
{\overline{\Lambda}} \equiv a^2 \beta{\overline{\lambda}}
   - \frac{\beta J}{S} \overline{\bf Q}\cdot\overline{\bf Q} a^2 ,\quad
{\tilde J} \equiv \beta J ,\quad
{\tilde B} \equiv \frac{\beta B}{2} .
\label{12d4}
\end{equation}
Evaluation of the Gaussian integrals yields
\begin{eqnarray}
Z_0 & \propto & e^{{\cal N} NS{\overline{\Lambda}}} \prod_{\bf k}
  \prod_{i\omega_n} \prod_\alpha
  \Bigg( -i\beta\hbar\omega_n + {\tilde J} S k^2a^2 \nonumber \\
& & {}- 2{\tilde J}\,\overline{\bf Q}\cdot{\bf k} a^2
  - {\tilde B} h_\alpha^\alpha + {\overline{\Lambda}}
  + \frac{{\tilde J}}{S} \overline{\bf Q}\cdot\overline{\bf Q} a^2
  \Bigg)^{\!-1} .
\end{eqnarray}
Writing the product as the exponential of a sum,
replacing the ${\bf k}$ sum by an integral,
$\sum_{\bf k} \to ({\cal N} a^2/4\pi^2) \int d^2k$,
and shifting ${\bf k}$ by $\overline{\bf Q}/S$, we obtain
\begin{eqnarray}
Z_0 & \propto & \exp\!\Bigg( {\cal N} NS{\overline{\Lambda}}
  - \frac{{\cal N} a^2}{4\pi^2} \int d^2k
    \sum_{i\omega_n} \sum_\alpha \nonumber \\
& & \times \ln \left[ -i\beta\hbar\omega_n
  + {\tilde J} S k^2a^2
  - {\tilde B} h_\alpha^\alpha + {\overline{\Lambda}} \right] \Bigg) .
\label{12Z5}
\end{eqnarray}
The MF partition function and thus all MF quantities only
depend on ${\overline{\lambda}}$ and $\overline{\bf Q}$ through
${\overline{\Lambda}}$. The saddle-point equation for
${\overline{\Lambda}}$ is
$\partial \ln Z_0/\partial{\overline{\Lambda}} = 0$, resulting in
\begin{eqnarray}
0 & = & {\cal N} NS - \frac{{\cal N} a^2}{4\pi^2} \int d^2k
  \sum_{i\omega_n} \sum_\alpha \nonumber \\
& & \times \frac1{-i\beta\hbar\omega_n + {\tilde J} S k^2a^2
    - {\tilde B} h_\alpha^\alpha + {\overline{\Lambda}}} .
\label{12x5}
\end{eqnarray}
The Matsubara sum in this expression is not well-defined since the
summands do not fall off fast enough. In writing it as a contour
integral the contribution from closing the contour does not vanish. The
usual procedure is to introduce a convergence factor
$e^{\pm i\eta\beta\hbar\omega_n}$ and let $\eta\to0$ afterwards. The
result is ambiguous, depending on the sign in the exponential. Here,
Eq.~(\ref{12x5}) only has solutions for positive sign. Consequently,
\begin{eqnarray}
0 & = & {\cal N} NS - \frac{{\cal N} a^2}{4\pi^2} \sum_\alpha
  \int d^2k\, n_B({\tilde J} S\,k^2a^2 - {\tilde B} h_\alpha^\alpha
    + {\overline{\Lambda}})   \nonumber \\
& = & {\cal N} NS + \frac{{\cal N}}{4\pi{\tilde J} S} \sum_\alpha
  \ln(1-e^{-{\overline{\Lambda}}+{\tilde B} h_\alpha^\alpha}) .
\label{12s5}
\end{eqnarray}
Here, $n_B(\epsilon)=1/(e^\epsilon-1)$ is the Bose function.
Eventually we find\cite{RS}
\begin{equation}
S = -\frac1{8\pi{\tilde J} S} \left[
  \ln(1-e^{-{\overline{\Lambda}}+{\tilde B}})
  + \ln(1-e^{-{\overline{\Lambda}}-{\tilde B}}) \right] .
\label{12s6}
\end{equation}
Equation (\ref{12s6}) for ${\overline{\Lambda}}$ can be evaluated
analytically. For given ${\overline{\Lambda}}$ we have the freedom to
choose $\overline{\bf Q}$,
and ${\overline{\lambda}}$ is then fixed by Eq.~(\ref{12d4}). This is a
consequence of gauge invariance since Eq.~(\ref{11g2}) specifies the
gauge only up to a constant. We choose $\overline{\bf Q} = 0$.
(The square lattice model without continuum approximation runs into
problems at this point since the quantity corresponding to
$\overline{\bf Q}$ shows a spurious first-order transition at the MF
level.)

The MF magnetization normalized so that $M_0(T=0) = S$ can be obtained
from Eq.~(\ref{12Z5})\cite{RS}
\begin{eqnarray}
M_0 & = & \frac{2}{{\cal N} N\beta}\,\frac{d}{dB} \ln Z_0 \nonumber \\
& = & -\frac1{8\pi{\tilde J} S} \left[
    \ln(1-e^{-{\overline{\Lambda}}+{\tilde B}})
  - \ln(1-e^{-{\overline{\Lambda}}-{\tilde B}}) \right] .
\label{12M6}
\end{eqnarray}
Some notes are in order: (i) Equation (\ref{12s6}) states that the
total number of ``up'' and ``down'' bosons (with $h_\alpha^\alpha=1$
and $-1$, respectively) is conserved, whereas Eq.~(\ref{12M6}) states
that the magnetization is basically the difference of the number of
``up'' and ``down'' bosons. (ii) The dependence of $Z_0$ on the field
$B$ through ${\overline{\Lambda}}$ is irrelevant at the MF level since
$\partial\ln Z_0/\partial{\overline{\Lambda}}=0$ by definition. This
is not the case at the $1/N$ level. (iii) The normalized magnetization
$M_0/S$ exhibits the universality mentioned in Sec.~\ref{intro}: It
only depends on ${\tilde J} S^2$ and ${\tilde B}$.\cite{RS}

Finally we compare the MF magnetization with the
original Heisenberg model. From Eqs.~(\ref{12s6}) and
(\ref{12M6}) we obtain at low temperatures
\begin{equation}
M_0 - S \cong \frac1{4\pi{\tilde J} S}\,\ln(1-e^{-\beta B})
\label{12M8}
\end{equation}
up to exponentially small corrections to the field $B$ of order of
${\overline{\Lambda}} - {\tilde B}
  \cong e^{-8\pi{\tilde J} S^2}/(1-e^{-\beta B})$. However,
Eq.~(\ref{12M8}) is just the magnetization of the Heisenberg model
neglecting magnon interactions. This means that the SU($N$)
MF theory captures the correct low-energy spin-wave
physics. Consequently, we expect higher
order corrections to be small for low $T$.

\subsection{$1/N$ corrections}
\label{2.3}

To take fluctuations in the auxiliary fields $\lambda$
and ${\bf Q}$ into account, we write
\begin{eqnarray}
\lambda({\bf r},\tau) & = & {\overline{\lambda}}
  + i\Delta\lambda({\bf r},\tau) ,
\label{12d1} \\
Q_j({\bf r},\tau) & = & 0 + \Delta Q_j({\bf r},\tau) .
\end{eqnarray}
The fluctuations in $\lambda$ are imaginary since $\lambda$
has to be integrated along the imaginary axis in Eq.~(\ref{11Z1}).
The fluctuations in $Q_j$ are real. They are subject to the
gauge constraint in Eq.~(\ref{11g2}).

We follow the procedure outlined by Auerbach.\cite{Auerbach}
The exact partition function is
\begin{equation}
Z = \int D\Delta\lambda\, D\Delta Q_j \exp(-N {\cal S}) ,
\label{13Z0}
\end{equation}
where the action ${\cal S}$ is expanded in a series for
small fluctuations $r_\ell$ with
$r_\ell$ standing for any mode $\Delta\lambda({\bf r},\tau)$ or
$\Delta Q_j({\bf r},\tau)$,
\begin{equation}
{\cal S} = \sum_{n=0}^\infty \frac1{n!}\,
  S_{\ell_1\ldots\ell_n}^{(n)}\,r_{\ell_1}\ldots r_{\ell_n} ,
\label{13A1}
\end{equation}
where summation over repeated field indices $\ell_i$ is here and in the
following implied. On the other hand, the action can be written as
${\cal S}={\cal S}_0+{\cal S}_{\text{dir}}+{\cal S}_{\text{loop}}$
with\cite{Auerbach}
\begin{eqnarray}
{\cal S}_0 & = & \frac1{N}\text{Tr}\ln G_0^{-1} ,  \\
{\cal S}_{\text{dir}} & = & \frac1{N\hbar} \int_0^{\hbar\beta} d\tau
  \int d^2r\, (NJ {\bf Q}\cdot{\bf Q}
  - NS \lambda) ,
\label{13A2} \\
{\cal S}_{\text{loop}} & = & \frac1{N}\text{Tr}\ln\!\left(
  1 + G_0 \upsilon_\ell r_\ell \right) ,
\end{eqnarray}
where the trace sums over space, time, and boson flavor,
$G_0$ is the MF bosonic Green function,
and $\upsilon_\ell$ is a vertex factor coupling
the fluctuation $r_\ell$ to two bosons.

\begin{figure}[htb]
\centerline{
\epsfig{file=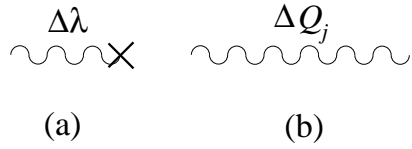,width=2.1in}}
\caption{Diagrams contributing to ${\cal S}_{\text{dir}}$.}
\label{fig131}
\end{figure}

The first term, ${\cal S}_0$, has the standard form for a
non-interacting system. It stems from the ${\bf k}$ integral part of
the MF free energy; see Eq.~(\ref{12Z5}).
The Green function can be read off from the MF partition function,
\begin{equation}
G_0^\alpha({\bf k},i\omega_n) = \left( -i\hbar\omega_n
  + J S k^2a^2 - B h_\alpha^\alpha/2 + a^2{\overline{\lambda}}
  \right)^{-1} .
\label{13G1}
\end{equation}
The second term, ${\cal S}_{\text{dir}}$, comes from the constant part
of the MF free energy but also contains fluctuations in the fields
$r_\ell$ which do not involve bosons. The constant part conspires with
${\cal S}_0$ to form the MF free energy $-\beta F_0 = N{\cal S}^{(0)}$.
The fluctuating part
contains a first order term in $\Delta\lambda$, corresponding to the
coupling of $\lambda$ to the constant $NS$ in Eq.~(\ref{11L4}),
and a second order term in $\Delta Q_j$ from the ${\bf Q}\cdot{\bf Q}$
term. The corresponding diagrams are shown in Fig.~\ref{fig131}.

\begin{figure}[htb]
\centerline{
\epsfig{file=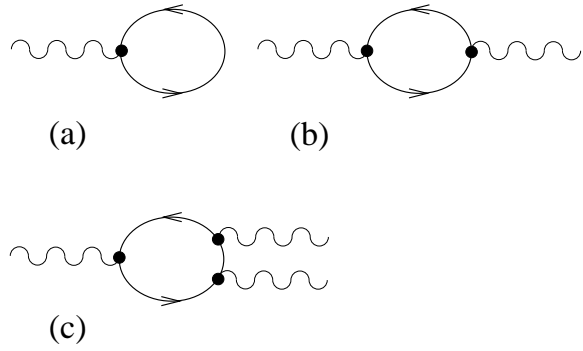,width=3in}}
\caption{Diagrams contributing to ${\cal S}_{\text{loop}}$.}
\label{fig132}
\end{figure}

The third term, ${\cal S}_{\text{loop}}$, contains the
contribution of fluctuations $r_\ell$ coupling to bosons. It is the
result of a linked-cluster expansion. By expanding the logarithm we
obtain the contribution from ${\cal S}_{\text{loop}}$ to
${\cal S}^{(n)}$,
\begin{equation}
\left.{\cal S}_{\ell_1\ldots\ell_n}^{(n)}\right|_{\text{loop}}
  = \frac1{N}\,\frac{(-1)^{n+1}}{n} \sum_{P_n}
  \text{Tr} \left(G_0 \upsilon_{\ell_1} \ldots
    G_0 \upsilon_{\ell_n}\right) .
\label{13A4}
\end{equation}
The sum $\sum_{P_n}$ runs over all permutations of the $n$ vertices.
The first few terms ${\cal S}^{(n)}$ are shown diagrammatically in
Fig.~\ref{fig132}. Solid lines with arrows denote MF boson Green
functions $G_0$ and the dots correspond to vertex factors
$\upsilon_\ell$. The wriggly lines are external legs $r_\ell$.
Disconnected diagrams are taken care of by a linked
cluster expansion, which puts the whole series into the exponential. No
internal $r_\ell$ lines appear since as far as the action is concerned
the $r_\ell({\bf r},\tau)$ are external variables.

For $n\ge 3$ Eq.~(\ref{13A4}) is the only contribution, whereas
${\cal S}^{(1)}$ and ${\cal S}^{(2)}$ contain contributions from
${\cal S}_{\text{dir}}$ and ${\cal S}_{\text{loop}}$.
The total first order term ${\cal S}^{(1)}$ can be shown to vanish as
it should since we are expanding around a saddle-point.

To find the vertex factors $\upsilon_\ell$ we write the exact partition
function $Z$ of Eq.~(\ref{11Z4}) in terms of Fourier transforms, where
the $b_\alpha$ dependent part of the Lagrangian is
\begin{eqnarray}
{\cal L}'' & = &
  \beta a^2 \sum_\alpha \left(
    - i\hbar\omega_n + JS k^2a^2 - \frac{B}2 h_\alpha^\alpha \right)
  \nonumber \\
& & \times b_\alpha^\ast({\bf k},i\omega_n)
  b_\alpha({\bf k},i\omega_n)
  + \frac{\beta a^4}{2\pi} \sum_\alpha \int d^2q \sum_{i\nu_n} 
  \nonumber \\
& & \times \left( -2J {\bf Q}({\bf q},i\nu_n)\cdot{\bf k} a^2
   + a^2 \lambda({\bf q},i\nu_n) \right) \nonumber \\
& & \times b_\alpha^\ast({\bf k},i\omega_n)
   b_\alpha({\bf k}-{\bf q},i\omega_n-i\nu_n) .
\label{13L2}
\end{eqnarray}
The first expression in parentheses is the inverse Green function
$(G_0^\alpha)^{-1}$. The same prefactors have to be included
in the vertex factors, which are the coefficients of the terms
$r_\ell b_\alpha^\ast b_\alpha$. Consequently,
\begin{eqnarray}
\upsilon_{\Delta\lambda} & = &
  \frac{a^2}{2\pi}\,\frac{4\pi^2}{{\cal N} a^2}\, i a^2
  = \frac{2\pi}{{\cal N}}\, i a^2 ,
\label{13v3} \\
\upsilon_{\Delta Q_j} & = &
  \frac{a^2}{2\pi}\,\frac{4\pi^2}{{\cal N} a^2}\, (-2J)\, a^2 k_j
  = -\frac{2\pi}{{\cal N}}\, 2J\,a^2 k_j .
\label{13v4}
\end{eqnarray}
The factor $4\pi^2/{\cal N} a^2$ in both cases stems from the integral
over ${\bf q}$. The factor of $i$ in $\upsilon_{\Delta\lambda}$ comes
from Eq.~(\ref{12d1}).

We now consider the expectation value
$\langle b_\alpha^\dagger b_\alpha\rangle$ for any $\alpha$
(no summation implied). From this we obtain two important quantities:
The average number of bosons per site
$\overline{n} = \sum_\alpha \langle b_\alpha^\dagger b_\alpha\rangle$,
and the magnetization
$M = N^{-1} \sum_\alpha h_\alpha^\alpha \langle b_\alpha^\dagger
  b_\alpha\rangle$.
Inserting a source term
$\Delta{\cal L}[j_\alpha] = \sum_\alpha j_\alpha b_\alpha^\ast b_\alpha$
into the Lagrangian (\ref{11L4}), where the source current
$j_\alpha$ is constant, we find
\begin{equation}
\langle b_\alpha^\dagger b_\alpha\rangle
  = -\frac1{{\cal N}\beta a^2}\,\frac1{Z}\,\left.
  \frac{\partial Z}{\partial j_\alpha} \right|_{j_\alpha=0} .
\end{equation}
Inserting the series expansion of Eq.~(\ref{13A1}), evaluating the
derivative, and expanding the exponential of the terms containing
${\cal S}^{(n)}$, $n\ge3$, we obtain
\begin{eqnarray}
\langle b_\alpha^\dagger b_\alpha\rangle & = &
  \frac{N}{{\cal N}\beta a^2Z} \int
  D\Delta\lambda\,D\Delta Q_j\, \left(\sum_{n=0}^\infty \frac1{n!}
    \frac{\partial {\cal S}_{\ell_1\ldots\ell_n}^{(n)}}
      {\partial j_\alpha}\,
    r_{\ell_1}\ldots r_{\ell_n} \right) \nonumber \\
& & \times \sum_{m=0}^\infty \frac{(-N)^m}{m!} \left(\sum_{n=3}^\infty
  \frac1{n!} {\cal S}_{\ell_1\ldots\ell_n}^{(n)} r_{\ell_1}\ldots
    r_{\ell_n} \right)^{\!\!m}
  \exp\!\left(-\frac{N}{2} {\cal S}_{\ell_1\ell_2}^{(2)} r_{\ell_1}
    r_{\ell_2}\right) .
\end{eqnarray}
All terms are Gaussian integrals, which can be evaluated by pairwise
contraction over the fields $r_\ell$. Diagrammatically,
any contraction is represented by connecting two vertices by
an {\it RPA fluctuation propagator} $D = ({\cal S}^{(2)})^{-1}$,
which we represent by a heavy wriggly line.

In the next step we calculate the $j_\alpha$ derivative of
${\cal S}^{(n)}$. The derivative
basically replaces $G_0$ by $-(G_0)^2$ so that we may expect it to
be related to ${\cal S}^{(n+1)}$. The Green function
in the presence of the source term is
$G_0^\alpha({\bf k},i\omega_n) = (-i\hbar\omega_n + J S k^2a^2
  - B h_\alpha^\alpha/2 + a^2{\overline{\lambda}} + a^2 j_\alpha)^{-1}$
so that its derivative is
$\partial G_0^\alpha/\partial j_\alpha = -G_0^\alpha a^2 G_0^\alpha$.
The vertex factor associated with $j_\alpha$ differs from
$\upsilon_{\Delta\lambda}$ only in a factor of $i$,
$\upsilon_{j_\alpha} = 2\pi{\cal N}^{-1}\, a^2$. With Eq.~(\ref{13A4})
we have
\begin{equation}
\left.\frac{\partial}{\partial j_\alpha}
  {\cal S}_{\ell_1\ldots\ell_n}^{(n)} \right|_{j_\alpha=0}
  = \frac{{\cal N}}{2\pi}\,\frac1{N}\,\frac{(-1)^{n+2}}{n} \sum_{P_n}
  \text{Tr} \left(G_0 \upsilon_{j_\alpha} G_0 \upsilon_{\ell_1} \ldots
    G_0 \upsilon_{\ell_n}
  + \ldots + G_0 \upsilon_{\ell_1} \ldots
    G_0 \upsilon_{j_\alpha} G_0 \upsilon_{\ell_n} \right) .
\end{equation}
The sum contains $nn!$ terms and not $(n+1)!$ because
$\upsilon_{j_\alpha}$ cannot appear to the right of $\upsilon_{\ell_n}$.
The invariance of the trace under cyclic rotation allows us to write
this expression as a sum over
all $(n+1)!$ permutations of the vertices $\upsilon_{j_\alpha}$,
$\upsilon_{\ell_1}, \ldots, \upsilon_{\ell_n}$, if we introduce a
correction factor for overcounting, $nn!/(n+1)! = n/(n+1)$.
We obtain\cite{Auerbach}
\begin{equation}
\left.\frac{\partial}{\partial j_\alpha}
  {\cal S}_{\ell_1\ldots\ell_n}^{(n)}
  \right|_{j_\alpha=0}
  = \frac{{\cal N}}{2\pi}\,\frac1{N}\,\frac{(-1)^{n+2}}{n+1}
  \sum_{P_{n+1}}
  \text{Tr} \left(G_0 \upsilon_{j_\alpha} G_0 \upsilon_{\ell_1} \ldots
    G_0 \upsilon_{\ell_n} \right)
  = \frac{{\cal N}}{2\pi}\,
    {\cal S}_{j_\alpha;\ell_1\ldots\ell_n}^{(n+1)} .
\end{equation}
Equation (17.25) in Ref.~\onlinecite{Auerbach} differs from this result
because of different definitions of vertex factors. It follows that
\begin{eqnarray}
\langle b_\alpha^\dagger b_\alpha\rangle & = &
  +\frac{N}{2\pi\beta a^2Z} \int
  D\Delta\lambda\,D\Delta Q_j\, \left(\sum_{n=0}^\infty \frac1{n!}
   {\cal S}_{j_\alpha;\ell_1\ldots\ell_n}^{(n+1)} r_{\ell_1}\ldots
   r_{\ell_n} \right) \nonumber \\
& & \times \sum_{m=0}^\infty \frac{(-N)^m}{m!} \left(\sum_{n=3}^\infty
  \frac1{n!} {\cal S}_{\ell_1\ldots\ell_n}^{(n)} r_{\ell_1}\ldots
    r_{\ell_n} \right)^{\!\!m}
  \exp\!\left(-\frac{N}{2} {\cal S}_{\ell_1\ell_2}^{(2)} r_{\ell_1}
    r_{\ell_2}\right) .
\label{13ex4}
\end{eqnarray}
In principle we can evaluate the integral for
any term in this series. The contraction of two variables gives
\begin{equation}
\frac1{Z} \int Dr_\ell\: r_{\ell_1} r_{\ell_2} \exp\!\left(-\frac{N}{2}
  {\cal S}_{\ell_1'\ell_2'}^{(2)} r_{\ell_1'} r_{\ell_2'}\right)
  = \frac1{N} ({\cal S}^{(2)})^{-1}_{\ell_1\ell_2} ,
\label{13Ga4}
\end{equation}
where the $r_\ell$ are assumed to be real.
(In fact they are real only in direct space but complex in Fourier
space. We can use the Gaussian integral for complex fields, which has
an additional factor of $2$, and note that
$\Delta\lambda(-{\bf q},-i\nu_n)$ and
$\Delta\lambda({\bf q},i\nu_n)$ are not independent since
$\Delta\lambda(-{\bf q},-i\nu_n) = \Delta\lambda^\ast({\bf q},i\nu_n)$
and similarly for $\Delta{\bf Q}$. Thus we have to restrict the
sum over ${\bf q}$ to one half space.
The factor of $1/2$ obtained in this way cancels the factor $2$ from
the Gaussian integral.)
The RPA propagator $D$ is the inverse of the matrix ${\cal S}^{(2)}$.
We obtain all terms in the expansion (\ref{13ex4}) by writing down
all allowed diagrams consisting of any number of boson loops with one
external $j_\alpha$ leg, represented by a dashed line,
and an even number of internal vertices, and connecting
the latter by RPA propagators in all possible ways. Allowed diagrams are
connected and do not contain loops with only one or two internal
vertices and no external vertex because first order terms cancel in an
expansion around a saddle point and the loop with two internal legs is
already included in the RPA propagator.

To figure out which terms are of which order in $1/N$, note that
the magnetization
$M = N^{-1} \sum_\alpha h_\alpha^\alpha \langle
  b_\alpha^\dagger b_\alpha\rangle$
contains an explicit factor of $1/N$. Furthermore, each loop
contributes a factor of $N$ from summation over flavors (the loop
with the external vertex does not contain a sum but has an explicit
$N$), and each RPA propagator contributes a factor of $1/N$;
see Eq.~(\ref{13Ga4}).

\begin{figure}[htb]
\centerline{
\epsfig{file=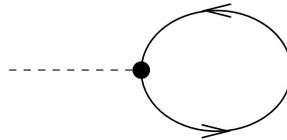,width=1.5in}}
\caption{The diagram for the MF magnetization.}
\label{fig132.3}
\end{figure}
\begin{figure}[htb]
\centerline{
\epsfig{file=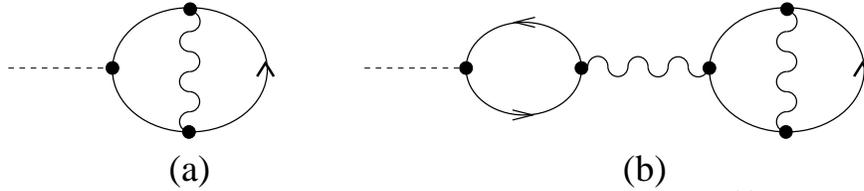,width=4.5in}}
\caption{$1/N$ diagrams contributing to the magnetization. Diagram (a)
corresponds to ${\cal D}_1^{(0)}$ in Eq.~(\protect\ref{13D1}) and (b)
corresponds to ${\cal D}_2^{(0)}$ in Eq.~(\protect\ref{13D2}).}
\label{fig132.5}
\end{figure}

The leading term is depicted in Fig.~\ref{fig132.3}. It is of order
$N^0$ in the
magnetization. This term reproduces the MF magnetization (\ref{12M6}).
Contributions of order $1/N$ in the magnetization have to contain
the same number of loops and propagators. The only two allowed diagrams
are shown in Fig.~\ref{fig132.5}. Similarly, we could write down
the diagrams to any order.

The two relevant contributions are, from Fig.~\ref{fig132.5}(a),
\begin{equation}
{\cal D}_1^{(0)} \equiv
\frac{N}{2\pi\beta a^2Z} \int Dr_\ell\,
  \frac12 {\cal S}_{j_\alpha;\ell_1\ell_2}^{(2+1)} r_{\ell_1} r_{\ell_2}
    \exp\!\left(-\frac{N}{2} {\cal S}_{\ell_1'\ell_2'}^{(2)}
    r_{\ell_1'} r_{\ell_2'}\right)
  = \frac{1}{4\pi\beta a^2} {\cal S}_{j_\alpha;\ell_1\ell_2}^{(2+1)}
    ({\cal S}^{(2)})^{-1}_{\ell_1\ell_2} ,
\label{13D1}
\end{equation}
and from Fig.~\ref{fig132.5}(b),
\begin{eqnarray}
{\cal D}_2^{(0)} & \equiv &
\frac{N}{2\pi\beta a^2Z} \int Dr_\ell\,
  {\cal S}_{j_\alpha;\ell_1}^{(1+1)} r_{\ell_1} \left(-\frac{N}{6}
  \right)
  {\cal S}_{\ell_2\ell_3\ell_4}^{(3)} r_{\ell_2} r_{\ell_3} r_{\ell_4}
    \exp\!\left(-\frac{N}{2} {\cal S}_{\ell_1'\ell_2'}^{(2)}
    r_{\ell_1'} r_{\ell_2'}\right)
  \nonumber \\
& = & -\frac{1}{4\pi\beta a^2} {\cal S}_{j_\alpha;\ell_1}^{(1+1)}
  {\cal S}_{\ell_2\ell_3\ell_4}^{(3)}
    ({\cal S}^{(2)})_{\ell_1\ell_2}^{-1}
    ({\cal S}^{(2)})_{\ell_3\ell_4}^{-1} .
\label{13D2}
\end{eqnarray}
${\cal D}_2^{(0)}$ contains a sum over all three possible pairings
of the four $r_\ell$. We have utilized the symmetry of ${\cal S}^{(3)}$
in each pair of indices. The evaluation of the relevant
${\cal S}^{(n)}$ is relegated to App.~\ref{appA}.
With the results found there we can write down the two terms in
$\langle b_\alpha^\dagger b_\alpha\rangle
  = {\cal D}_1^{(0)} + {\cal D}_2^{(0)}$,
\begin{eqnarray}
{\cal D}_1^{(0)} & = & \frac1{2{\cal N}} \sum_{{\bf q},i\nu_n} \left(
  \frac{\sigma_0^{\alpha\prime}}{\sum_\beta \sigma_0^\beta} +
  \frac{\sigma_\perp^{\alpha\prime}}{{\cal N} N/2{\tilde J}+\sum_\beta
  \sigma_\perp^\beta} \right) , \\
{\cal D}_2^{(0)} & = & -\frac{1}{{\cal N} N}\,
  \frac{n_B({\overline{\Lambda}}-{\tilde B} h_\alpha^\alpha)}
  {n_B({\overline{\Lambda}}-{\tilde B})+n_B({\overline{\Lambda}}
    +{\tilde B})} \sum_{{\bf q},i\nu_n} \left(
  \frac{\sum_\beta \sigma_0^{\beta\prime}}{\sum_\beta \sigma_0^\beta} +
  \frac{\sum_\beta \sigma_\perp^{\beta\prime}}
    {{\cal N} N/2{\tilde J}+\sum_\beta \sigma_\perp^\beta} \right) ,
\end{eqnarray}
where we have introduced new symbols,
\begin{eqnarray}
\sigma_0^\alpha & \equiv & \sum_{\bf k} \frac{n_B(\epsilon_{{\bf k}
    +{\bf q}/2}^\alpha)
  -n_B(\epsilon_{{\bf k}-{\bf q}/2}^\alpha)}
    {-i\beta\hbar\nu_n+2{\tilde J} S\, q k_1 a^2} ,
\label{13ds1} \\
\sigma_\perp^\alpha & \equiv & \sum_{\bf k} k_2^2a^2
  \frac{n_B(\epsilon_{{\bf k}+{\bf q}/2}^\alpha)
  -n_B(\epsilon_{{\bf k}-{\bf q}/2}^\alpha)}{-i\beta\hbar\nu_n
    +2{\tilde J} S\, q k_1 a^2} , \\
\sigma_0^{\alpha\prime} & \equiv & \sum_k \frac{n_B^{(1)}(
  \epsilon_{{\bf k}+{\bf q}/2}^\alpha)
  -n_B^{(1)}(\epsilon_{{\bf k}-{\bf q}/2}^\alpha)}
  {-i\beta\hbar\nu_n+2{\tilde J} S\, q k_1 a^2} , \\
\sigma_\perp^{\alpha\prime} & \equiv & \sum_k  k_2^2a^2
  \frac{n_B^{(1)}(\epsilon_{{\bf k}+{\bf q}/2}^\alpha)
  -n_B^{(1)}(\epsilon_{{\bf k}-{\bf q}/2}^\alpha)}
  {-i\beta\hbar\nu_n+2{\tilde J} S\, q k_1 a^2}
\end{eqnarray}
with $\epsilon_{\bf k}^\alpha \equiv {\tilde J} S\,k^2a^2
- {\tilde B} h_\alpha^\alpha + {\overline{\Lambda}}$
and the derivative $n_B^{(1)}$ of the Bose function.

Before we turn to the magnetization we look at the total number of
bosons per site
$\overline{n} = \sum_\alpha \langle b_\alpha^\dagger b_\alpha\rangle$.
The MF contribution is $\overline{n}_0 = NS$ since
${\overline{\Lambda}}$ was chosen that way. To the next order,
\begin{eqnarray}
\overline{n} & = & NS + \frac1{2{\cal N}} \sum_{{\bf q},i\nu_n} \left(
  \frac{\sum_\alpha \sigma_0^{\alpha\prime}}{\sum_\beta \sigma_0^\beta}
  + \frac{\sum_\alpha \sigma_\perp^{\alpha\prime}}
    {{\cal N} N/2{\tilde J}+\sum_\beta \sigma_\perp^\beta} \right)
  \nonumber \\
& & {}- \frac1{{\cal N} N}\,\frac{N}{2}\,\frac{n_B({\overline{\Lambda}}
  -{\tilde B})+n_B({\overline{\Lambda}}+{\tilde B})}
  {n_B({\overline{\Lambda}}-{\tilde B})
  +n_B({\overline{\Lambda}}+{\tilde B})} \sum_{{\bf q},i\nu_n} \left(
  \frac{\sum_\beta \sigma_0^{\beta\prime}}{\sum_\beta \sigma_0^\beta} +
  \frac{\sum_\beta \sigma_\perp^{\beta\prime}}
    {{\cal N} N/2{\tilde J}+\sum_\beta \sigma_\perp^\beta} \right)
  = NS + 0 .
\end{eqnarray}
We thus see explicitly that the constraint (\ref{11c6}) is still
satisfied at the $1/N$ level. This is a special case of Auerbach's
general proof\cite{Auerbach} that the constraint is satisfied to any
order. Thus we need not ``shift the saddle point,'' {\it i.e.}, adjust
${\overline{\Lambda}}$ so that the constraint is still satisfied.

We now calculate the $1/N$ contribution to the magnetization,
\begin{equation}
M = M_0 - \frac1{2{\cal N} N} \sum_{{\bf q},i\nu_n} \left(
  \frac{\sum_\alpha (c_1-h_\alpha^\alpha) \sigma_0^{\alpha\prime}}
    {\sum_\beta \sigma_0^\beta} +
  \frac{\sum_\alpha (c_1-h_\alpha^\alpha) \sigma_\perp^{\alpha\prime}}
    {{\cal N} N/2{\tilde J}+\sum_\beta \sigma_\perp^\beta} \right) ,
\end{equation}
where $M_0$ is the MF magnetization (\ref{12M6}) and
\begin{equation}
c_1 \equiv \frac{n_B({\overline{\Lambda}}-{\tilde B})
  -n_B({\overline{\Lambda}}+{\tilde B})}{n_B({\overline{\Lambda}}
    -{\tilde B})+n_B({\overline{\Lambda}}+{\tilde B})} .
\label{13c8}
\end{equation}
Expressing the momentum sum by an integral we find
\begin{eqnarray}
M & = & M_0 - \frac1{N}\,\frac{a^2}{8\pi^2} \int d^2q \sum_{i\nu_n}
  \left( \frac{\sum_\alpha (c_1-h_\alpha^\alpha)
   \sigma_0^{\alpha\prime}}{\sum_\beta \sigma_0^\beta} +
   \frac{\sum_\alpha (c_1-h_\alpha^\alpha) \sigma_\perp^{\alpha\prime}}
     {{\cal N} N/2{\tilde J}+\sum_\beta \sigma_\perp^\beta} \right)
  \nonumber \\
& \equiv & M_0 - \frac1{N}\,\frac{a^2}{8\pi^2} \int d^2q \sum_{i\nu_n}
    \big( \Delta M_0({\bf q},i\nu_n) + \Delta M_\perp({\bf q},i\nu_n)
  \big) .
\label{13M8}
\end{eqnarray}
The first term in the parentheses is due to fluctuations
$\Delta\lambda$ while the second comes from $\Delta{\bf Q}$.

The same result (\ref{13M8}) can be found by writing down the
one-loop contribution to the
free energy and taking the derivative with respect to magnetic field.
In this way we find physical interpretations for the two $1/N$ diagrams:
The term coming from the explicit $B$ dependence of the free energy
corresponds to Fig.~\ref{fig132.5}(a), whereas the indirect dependence
through ${\overline{\Lambda}}(B)$ corresponds to Fig.~\ref{fig132.5}(b).

Although the expression (\ref{13M8}) for the magnetization is formally
correct, great care is needed in evaluating the frequency sum. We will
first show briefly how naive evaluation leads to a spurious divergence
of the integral over momentum ${\bf q}$ and then present the solution
to this problem. The solution involves carefully taking into account
normal ordering of boson operators.
First, we derive the contributions to the magnetization for
large momentum ${\bf q}$ of the (in Fig.~\ref{fig132.5} vertical) RPA
propagator. We express the momentum sum in
$\sigma_0^\alpha$ and $\sigma_\perp^\alpha$ by an integral and shift
the variable ${\bf k}$
in the two summands by ${\bf k}\to{\bf k}\mp{\bf q}/2$,
\begin{eqnarray}
\sigma_{0,\perp}^\alpha\!\!\! & = & \!\frac{{\cal N} a^2}{4\pi^2}
  \int d^2k
  \left\{\!{1 \atop k_2^2 a^2}\!\right\}
  \left( \frac1{-i\beta\hbar\nu_n
    +2{\tilde J} S\, q k_1 a^2-{\tilde J} S q^2a^2}
    - \frac1{-i\beta\hbar\nu_n+2{\tilde J} S\, q k_1 a^2
    +{\tilde J} S q^2a^2} \right)  \nonumber \\
& & \times n_B({\tilde J} S k^2a^2-{\tilde B} h_\alpha^\alpha
  +{\overline{\Lambda}}) ,
\end{eqnarray}
where the upper (lower) expression in $\{\}$ pertains to
$\sigma_0^\alpha$ ($\sigma_\perp^\alpha$), and
$\sigma_0^{\alpha\prime}$ and $\sigma_\perp^{\alpha\prime}$ are
obtained by replacing $n_B$ by $n_B^{(1)}$.
Formally, we expand for small $k_1$
since large ${\bf k}$ are exponentially suppressed.
Odd powers of $k_1$ vanish so that
\begin{eqnarray}
\sigma_{0,\perp}^\alpha & = & \frac{{\cal N} a^2}{4\pi^2}
  \sum_{m\;\text{even}}
  \left( \frac1{(-i\beta\hbar\nu_n-{\tilde J} S\,q^2a^2)^{m+1}}
       - \frac1{(-i\beta\hbar\nu_n+{\tilde J} S\,q^2a^2)^{m+1}} \right)
  \nonumber \\
& & \times \int d^2k \left\{{1\atop k_2^2a^2}\right\} (2{\tilde J}
  S\,qk_1 a^2)^m n_B({\tilde J} S\,k^2a^2 - {\tilde B} h_\alpha^\alpha
    + {\overline{\Lambda}}) .
\label{13d8}
\end{eqnarray}
This is an assymptotic series and does not converge. However, for
our argument we only need the first two non-vanishing terms, which are
well-defined.

The leading term in $\sum_\alpha \sigma_0^\alpha$ reads
\begin{eqnarray}
\sum_\alpha \sigma_0^\alpha
& \cong & - \frac{{\cal N}}{4\pi{\tilde J} S}
  \left( \frac1{-i\beta\hbar\nu_n-{\tilde J} S\,q^2a^2}
       - \frac1{-i\beta\hbar\nu_n+{\tilde J} S\,q^2a^2} \right)
  \sum_\alpha \ln(1-e^{-{\overline{\Lambda}}
    +{\tilde B} h_\alpha^\alpha})  \nonumber \\
& = & {\cal N} NS \left( \frac1{-i\beta\hbar\nu_n-{\tilde J} S\,q^2a^2}
  - \frac1{-i\beta\hbar\nu_n+{\tilde J} S\,q^2a^2} \right) ,
\end{eqnarray}
where the last step follows from the saddle-point equation
(\ref{12s6}). Thus the
$m=0$ term is independent of magnetic field. Similarly we find
$\sum_\alpha (c_1 - h_\alpha^\alpha) \sigma_0^{\alpha\prime} \cong 0$
to the same order, where we have used Eq.~(\ref{13c8}).
This result is not surprising since
$\sum_\alpha (c_1 - h_\alpha^\alpha) \sigma_0^{\alpha\prime}$ is, up to
a factor, the magnetic field derivative of
$\sum_\alpha \sigma_0^\alpha$. The leading non-vanishing ($m=2$) term is
\begin{eqnarray}
\sum_\alpha (c_1 - h_\alpha^\alpha) \sigma_0^{\alpha\prime}
& \cong & \frac{{\cal N} a^2}{4\pi^2}
  \left( \frac1{(-i\beta\hbar\nu_n-{\tilde J} S\,q^2a^2)^3}
    - \frac1{(-i\beta\hbar\nu_n+{\tilde J} S\,q^2a^2)^3} \right)
  \nonumber \\
& & \times \int d^2k\, (2{\tilde J} S\,qk_1 a^2)^2
  \sum_\alpha (c_1 - h_\alpha^\alpha)\,
  n_B^{(1)}({\tilde J} S\,k^2a^2 - {\tilde B} h_\alpha^\alpha
    + {\overline{\Lambda}})
\end{eqnarray}
and consequently, for large ${\bf q}$,
\begin{eqnarray}
\Delta M_0({\bf q},i\nu_n) =
\frac{\sum_\alpha (c_1-h_\alpha^\alpha)
  \sigma_0^{\alpha\prime}}{\sum_\beta \sigma_0^\beta}
& \cong & \frac{{\tilde J} a^4}{2\pi^2 N}
  \left( \frac1{-i\beta\hbar\nu_n-{\tilde J} S\,q^2a^2}
    - \frac1{-i\beta\hbar\nu_n+{\tilde J} S\,q^2a^2} \right)
  \nonumber \\
& & \times \sum_\alpha (c_1-h_\alpha^\alpha) \int d^2k\,k_1^2\,
  n_B^{(1)}({\tilde J} S\,k^2a^2 - {\tilde B} h_\alpha^\alpha
    + {\overline{\Lambda}}) .
\label{13Si0.8}
\end{eqnarray}
Adding the two fractions under the sum and performing the Matsubara sum
we find
\begin{equation}
\sum_{i\nu_n} \left( \frac1{-i\beta\hbar\nu_n-{\tilde J} S\,q^2a^2}
    - \frac1{-i\beta\hbar\nu_n+{\tilde J} S\,q^2a^2} \right)
  = \sum_{i\nu_n} \frac{-2{\tilde J} S\,q^2a^2}{(\beta\hbar\nu_n)^2
    +({\tilde J} S\,q^2a^2)^2}
  = - \coth \frac{{\tilde J} S\,q^2a^2}{2} .
\label{13lq6}
\end{equation}
For large momentum ${\bf q}$ we thus obtain
\begin{equation}
\sum_{i\nu_n} \Delta M_0({\bf q},i\nu_n)
  \cong -\frac1{4\pi N {\tilde J} S^2} \sum_\alpha (c_1-h_\alpha^\alpha)
  \ln(1-e^{-{\overline{\Lambda}}+{\tilde B} h_\alpha^\alpha})
  = c_1 - \frac{M_0}{S} .
\label{13c10}
\end{equation}
The large momentum limit of the $\Delta{\bf Q}$ contribution is found
more easily. We have to consider $\Delta M_\perp({\bf q},i\nu_n)$
for large momenta; see Eq.~(\ref{13M8}).
The leading order ($m=0$) term of $\sum_\beta \sigma_\perp^\beta$
is negligible compared to the constant added to it. Furthermore, the
$m=0$ term in
$\sum_\alpha (c_1-h_\alpha^\alpha) \sigma_\perp^{\alpha\prime}$ does not
vanish. Thus the leading term is
\begin{equation}
\frac{{\tilde J} a^4}{2\pi^2 N} \left( \frac1{-i\beta\hbar\nu_n
    -{\tilde J} S\,q^2a^2}
  - \frac1{-i\beta\hbar\nu_n+{\tilde J} S\,q^2a^2} \right)
    \sum_\alpha (c_1-h_\alpha^\alpha) \int d^2k\,k_2^2\,
    n_B^{(1)}({\tilde J} S\,k^2a^2 - {\tilde B} h_\alpha^\alpha
      + {\overline{\Lambda}}) .
\end{equation}
This expression is equal to Eq.~(\ref{13Si0.8}). Consequently, for
large momenta the integrand of the external momentum integral in
Eq.~(\ref{13M8}) is
\begin{equation}
2\left( c_1 - M_0/S \right) ,
\label{13lq8}
\end{equation}
which is independent of momentum. This term would lead to a strong
UV divergence of the ${\bf q}$ integral in Eq.~(\ref{13M8}).

We now discuss the cure. As mentioned above,
we have to normal order the operators before we can write the partition
function as a functional integral. Careful
treatment of ordering is often essential to resolve ambiguities in
the expectation value of operator products at equal times.
Here we are interested in
$\langle b_\alpha^\dagger({\bf r},\tau) b_\alpha({\bf r},\tau)\rangle$.
One way of dealing with these ambiguities is to split the time of the
operators in the Lagrangian in such a way that the creation operator
is at an infinitesimally later time than the
annihilation operator. The time-ordered product in the definition of
the Green function then takes care of normal ordering at equal times.

If the action contains a term like
\begin{equation}
\Delta{\cal S} = \frac1{N\hbar} \int_0^{\hbar\beta} d\tau \int d^2r
  \sum_\alpha
  c_\alpha({\bf r},\tau)\, b_\alpha^\ast({\bf r},\tau+\eta)
  b_\alpha({\bf r},\tau) ,
\end{equation}
where $\eta>0$ is small, then, after Fourier transformation,
\begin{equation}
\Delta{\cal S} = \frac{\beta a^6}{2\pi N} \int d^2k\,d^2q
  \sum_{i\omega_n,i\nu_n} \sum_\alpha
  c_\alpha({\bf q},i\nu_n)\,e^{i\omega_n\eta}\,
  b_\alpha^\ast({\bf k},i\omega_n)
  b_\alpha({\bf k}-{\bf q},i\omega_n-i\nu_n) .
\end{equation}
The coefficient $c_\alpha$ obtains a phase factor
$e^{i\omega_n\eta}$, where $i\omega_n$ is the frequency of the boson
created at this point. We split the time in this way in the exact
Lagrangian (\ref{11L4}) as well as in the source term
$j_\alpha b_\alpha^\ast b_\alpha$. As a consequence, phase factors
containing the frequency of the outgoing boson appear at all vertices.
The only places where they are relevant turn out to be in
${\cal S}^{(3)}$ and ${\cal S}^{(2+1)}$.
In the first term in Eq.~(\ref{13A6}) we obtain a total factor of
$e^{i\omega_n\eta} e^{i(\omega_n+\nu_n)\eta} e^{i\omega_n\eta}
  = e^{3i\omega_n\eta} e^{i\nu_n\eta}$
from the three vertices. The factor containing $i\omega_n$
is irrelevant since the sum over $i\omega_n$ is unambiguous anyway.
We are left with an overall factor
of $e^{i\nu_n\eta}$. The second term from the
symmetrization in Eq.~(\ref{13A6})
obtains a factor $e^{-i\nu_n\eta}$. If the two terms are added,
the terms which have the denominators squared obtain a prefactor of
$2i\sin \nu_n\eta$, which vanishes in the limit $\eta\to0$
(the denominators are already of second order in frequency so that
ambiguities or divergences do not arise) and the remaining terms are
\begin{eqnarray}
{\cal S}_{\Delta\lambda(0,0),\Delta\lambda(-{\bf q},-i\nu_n),
  \Delta\lambda({\bf q},i\nu_n)}^{(3)}
& = & \frac{(2\pi)^3}{{\cal N}^3 N}\,i\beta a^2 \sum_{\bf k} \sum_\alpha
  \frac{e^{-i\nu_n\eta} n_B^{(1)}(\epsilon_{{\bf k}+{\bf q}/2}^\alpha) -
    e^{i\nu_n\eta} n_B^{(1)}(\epsilon_{{\bf k}-{\bf q}/2}^\alpha)}
    {-i\beta\hbar\nu_n+2{\tilde J} S\,q k_1 a^2} (-\beta^2 a^4) ,
\label{13A17c} \\
{\cal S}_{\Delta\lambda(0,0),\Delta Q_2(-{\bf q},-i\nu_n),
  \Delta Q_2({\bf q},i\nu_n)}^{(3)}
& = & \frac{(2\pi)^3}{{\cal N}^3 N}\,i\beta a^2 \sum_{\bf k} \sum_\alpha
  \frac{e^{-i\nu_n\eta} n_B^{(1)}(\epsilon_{{\bf k}+{\bf q}/2}^\alpha) -
    e^{i\nu_n\eta} n_B^{(1)}(\epsilon_{{\bf k}-{\bf q}/2}^\alpha)}
    {-i\beta\hbar\nu_n+2{\tilde J} S\,q k_1 a^2} 4{\tilde J}^2 a^4 k_2^2
    .
\label{13A17d}
\end{eqnarray}
${\cal S}^{(2+1)}$ follows as above.

The Matsubara sum of the leading large ${\bf q}$ term is now, instead
of Eq.~(\ref{13lq6}),
\begin{eqnarray}
\lefteqn{
\sum_{i\nu_n} \left( \frac{e^{-i\nu_n\eta}}
    {-i\beta\hbar\nu_n-{\tilde J} S\,q^2a^2}
  - \frac{e^{+i\nu_n\eta}}{-i\beta\hbar\nu_n+{\tilde J} S\,q^2a^2}
  \right)_{\eta\to0^+} }  \nonumber \\
& & \qquad = \sum_{i\nu_n} \left( \frac{-2{\tilde J} S\,q^2a^2}
    {(\beta\hbar\nu_n)^2+({\tilde J} S\,q^2a^2)^2}
  \cos \nu_n\eta + \frac{2\beta\hbar\nu_n}
    {(\beta\hbar\nu_n)^2+({\tilde J} S\,q^2a^2)^2}
  \sin \nu_n\eta \right)_{\eta\to0^+} \nonumber \\
& & \qquad = - \coth \frac{{\tilde J} S\,q^2a^2}{2} + \lim_{\eta\to0^+}
  \frac{\sinh {\tilde J} S\,q^2a^2 (1/2-\eta)}
  {\sinh {\tilde J} S\,q^2a^2/2} = 1 - \coth \frac{{\tilde J} S\,q^2a^2}
    {2} .
\label{13lq16}
\end{eqnarray}
The sine series exactly cancels the leading term of the naive series
for large ${\bf q}$, making the remaining expression exponentially
small.
In other words, the phase factors introduced to ensure correct operator
ordering just remove the constant (\ref{13lq8}) from the integrand for
large ${\bf q}$. The exponential factors are irrelevant in all
other terms, which are of higher order in $1/i\nu_n$ and are thus
unambiguous.

Deriving a convergence factor $e^{\pm i\nu_n\eta}$
is a common method to resolve the ambiguity of a Matsubara sum.
What is unusual here is that two different
factors appear for the two terms. It is easy to fall into the
trap of thinking that one should simply add the two fractions under
the frequency sum in Eq.~(\ref{13lq6}),
arguing that the sum then looks unambiguous. This loses an essential
contribution because of the two different phase factors.

\begin{figure}[htb]
\centerline{
\epsfig{file=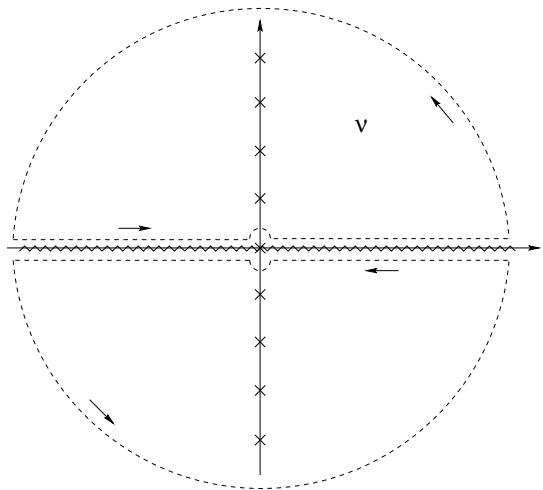,width=2.8in}}
\caption{Contour of integration for the Matsubara sum over $i\nu_n$.}
\label{fig134}
\end{figure}

We utilize the above result by calculating the ${\bf q}$ integrand
numerically without taking normal ordering into account and then
subtracting the constant (\ref{13lq8}) explicitly.
The frequency sum is expressed in terms of a contour integral. First we
analyze the analytic structure of $\sigma_0^\alpha$ etc.\ in the
complex $\nu$ plane. If we replace the sum over ${\bf k}$ by an
integral in Eq.~(\ref{13ds1}) for $\sigma_0^\alpha$ we see that
the integrand of the $k_1$ integral has a pole and, consequently,
$\sigma_0^\alpha$ has a branch cut.
Furthermore, it can be shown
that $\sum_\alpha \sigma_0^\alpha$ does not have zeros
apart from the trivial case ${\bf q}=0$.
The quantity $\sum_\alpha (c_1-h_\alpha^\alpha) \sigma_0^{\alpha\prime}$
obviously also has a branch cut along the real axis and
does not have poles. Consequently, the $\Delta\lambda$
contribution to the magnetization, $\Delta M_0$,
has a branch cut and no poles.
The contour of integration, ${\cal C}$, is shown in Fig.~\ref{fig134}.
We have
\begin{eqnarray}
\lefteqn{
\frac1{\hbar\beta} \sum_{i\nu_n\neq 0} \Delta M_0
  = \frac1{2\pi i} \oint_{\cal C} d\nu\, n_B(\beta\hbar\nu)\,
  \Delta M_0 } \nonumber \\
& & = \frac1{2\pi i} \bigg[
  - \int_{{\cal C}_\epsilon}\! d\nu\, n_B(\beta\hbar\nu)\,
      \Delta M_0(0) \nonumber \\
& & \qquad{}+ \left( \int_{-\infty}^{-\epsilon}\!\!
    + \int_\epsilon^\infty \right)\! d\nu\,
  n_B(\beta\hbar\nu)\, [-i A_0(\nu)] \bigg],
\end{eqnarray}
where $A_0(\nu) \equiv -2\,\text{Im}\, \Delta M_0(\nu+i\delta)$ is the
spectral function of $\Delta M_0$ and ${\cal C}_\epsilon$ is a
positively directed circular path of radius $\epsilon\to0$ around the
origin. Since $A_0$ vanishes continuously at the origin we get, after
moving the last term to the left hand side,
\begin{equation}
\frac1{\hbar\beta} \sum_{i\nu_n} \Delta M_0
  = -\frac1{2\pi} \int_{-\infty}^\infty d\nu\,n_B(\beta\hbar\nu)\,
    A_0(\nu) .
\end{equation}
A similar equation holds for the $\Delta{\bf Q}$ contribution,
$\Delta M_\perp$. We do not discuss the numerical methods in any detail
since they are standard. We only note that it is useful to expand the
Bose functions $n_B$ in $\sigma_0^\alpha$ etc.\
in a geometric series because this allows one to perform the integrals
over ${\bf k}$ analytically, thereby replacing 2D integrals
by numerical summation of well-behaved series.

After evaluating the frequency sum, we have to subtract the constant
(\ref{13lq8}). Numerically the correction term is indeed found to
cancel the constant for large ${\bf q}$.
The new leading term drops off as $1/q^2$ so that
the ${\bf q}$ integral diverges only logarithmically. We regularize the
integral by restricting it to the first Brillouin zone, {\it i.e.}, by
a lattice cutoff. We use a circular Brillouin zone. The integration
over the angle of ${\bf q}$ is then trivial.

We find that fluctuations in $\lambda$ and ${\bf Q}$ always decrease the
magnetization, as is intuitively expected. In fact
the magnetization to order $1/N$ can become slightly negative.
Of course, the exact magnetization cannot be negative. Apparently
the $1/N$ expansion does not work well for SU(2).
We can force the result to be positive by putting the fluctuations
into an exponential: In writing down the functional integral we should
impose the constraint that the total magnetization be positive. This
constraint can be implemented by writing the full magnetization as
$M = M_0 e^g$ and expanding $g$ in powers of $1/N$. If
$M = M_0 + M_{1/N} + {\cal O}(1/N)^2$ then
$g = 1 + M_{1/N}/M_0 + {\cal O}(1/N)^2$.
Both expansions are {\it equally valid\/} to order $1/N$.
Of course, this method seems dubious if the $1/N$ term is not small.

\begin{figure}[tb]
\centerline{
\epsfig{file=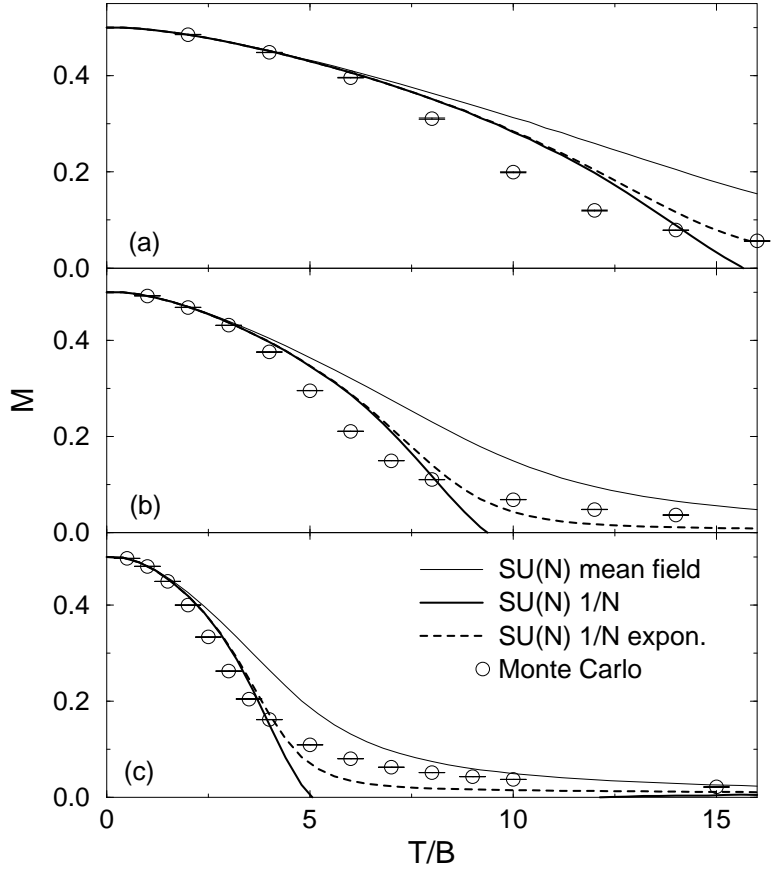,width=4in}}
\caption{SU($N$) magnetization for magnetic fields (a) $B/J=0.05$,
(b) $B/J=0.1$, and (c) $B/J=0.25$ as a function of temperature in units
of $B$. The thin solid line is the MF magnetization, the thick solid
line includes $1/N$ corrections linearly, the thick dashed line
includes them in the exponential to force positivity, and the circles
with error bars are quantum Monte Carlo results for a $32\times32$
lattice.\protect\cite{letter,PH}}
\label{fig135}
\end{figure}

In Fig.~\ref{fig135} we plot the magnetization for $S=1/2$ and the
fields $B/J=0.05, 0.1, 0.25$ as a function of $T/B$.
At the lowest temperatures the SU($N$) model describes the Monte Carlo
results quite well, as expected since the SU($N$)
model captures the correct low-energy physics. The $1/N$ corrections
are thus very small here. At high temperatures,
however, the $1/N$ term is too large for all considered fields.
Although the results with the exponentiation trick are better and show
the correct qualitative behavior, they are not quantitatively
better than the MF results.
We discuss the results further at the end of the next section.

\section{O($N$) model}

\subsection{General considerations}
\label{3.1}

In the last section the Heisenberg model was rewritten in terms of Bose
fields and the resulting SU(2) model with two boson flavors was
generalized to SU($N$). The homomorphism between the groups SU(2) and
O(3) opens another
way to obtain a large $N$ theory. In this section the Heisenberg
model is mapped onto an O(3) model, which is then generalized
to O($N$). Many concepts are identical to the SU($N$) case.
However, the O($N$) model adds a number of complications.

The group theory involved here can be found else\-where.\cite{groups}
In brief, the Lie groups SU(2) and O(3) have the same algebra, up to
different representations. This means that the infinitesimal generators
of SU(2), namely the Pauli matrices, have the same commutation
relations as the three infinitesimal generators of O(3),
$(X_k)_{ij} = -2i \epsilon_{ijk}$. Put informally, SU(2)
and O(3) have the same local structure, although the global structure
is different. Note that we could also talk about the SO(3) model instead
of O(3) since the two have the same algebra.

The upshot of this is that we can
map the Heisenberg model onto an O(3) model. We introduce three
Bose fields $b_\alpha$ and let
$S^k = -i \epsilon_{k\alpha\beta} b_\alpha^\dagger b_\beta$, $k=x,y,z$,
where we again assume summation over repeated indices. It is easily
shown that the commutators of the spin components $S^k$ are correct.
To restrict the Hilbert space to the physical states
two constraints are needed,
$b_\alpha^\dagger b_\alpha = S$ and
$b_\alpha^\dagger b_\alpha^\dagger = 0$.
The second constraint needs explanation.
Let us consider the eigenstates of $S^z$ for a single spin.
Since $S^z$ is not diagonal in boson flavors we introduce new bosons
$c_1 \equiv (b_1+ib_2)/\sqrt{2}$,
$c_2 \equiv (b_1-ib_2)/\sqrt{2}$, and $c_3 \equiv b_3$.
Then $S^z = -c_1^\dagger c_1 + c_2^\dagger c_2$ and the constraints read
$c_\alpha^\dagger c_\alpha = S$ and
$c_1^\dagger c_2^\dagger + c_2^\dagger c_1^\dagger
  + c_3^\dagger c_3^\dagger = 0$.
The eigenstates of $S^z$ are simultaneous eigenstates to
$c_1^\dagger c_1$, $c_2^\dagger c_2$, and $c_3^\dagger c_3$. As an
example, the following table shows the eigenvalues
of the number operators and of $S^z$ for $S=2$. The first constraint
means that there are two bosons.
\begin{displaymath}
\begin{array}{ccc|c}
  c_1^\dagger c_1 & c_2^\dagger c_2 & c_3^\dagger c_3 & S^z \\ \hline
  2 & 0 & 0 & -2 \\
  1 & 0 & 1 & -1 \\
  1 & 1 & 0 & 0  \\
  0 & 1 & 1 & 1  \\
  0 & 2 & 0 & 2  \\
  0 & 0 & 2 & 0
\end{array}
\end{displaymath}
The state with $S^z=0$ is obviously counted twice. The second
constraint just removes the last state. It can be rewritten as
$c_3^\dagger c_3^\dagger |\psi\rangle
  = -2c_1^\dagger c_2^\dagger |\psi\rangle$,
where $|\psi\rangle$ is any state. This means that the state one gets
by creating two $c_3$ bosons is the same as the one
produced by creating one $c_1$ and one $c_2$. The second constraint thus
reduces the Hilbert space by identifying states with one another.
For general $S$, the second constraint removes spurious
spin multiplets of lower total spin.

The first constraint does not make sense for half integer spin. We
assume $S$ integer. We will not have to do this for even $N$ in
O($N$) theory.

The O(3) spin matrix should be an element of the algebra, which
consists of antisymmetric $3\times3$ matrices. In three dimensions
any antisymmetric matrix is dual to an axial vector. This is in fact
the reason why angular momenta can be written as axial vectors in
three dimensions. Here
we go the opposite way and define the spin matrix by
$S_\beta^\alpha \equiv i \epsilon_{\alpha\beta k} S^k
  = b_\alpha^\dagger b_\beta - b_\beta^\dagger b_\alpha$.
Using the antisymmetry of $S_\beta^\alpha$, the Hamiltonian
(\ref{11H1}) becomes
\begin{equation}
H = -\frac{J}2 \sum_{\langle ij\rangle} S_\beta^\alpha(i)
  S_\alpha^\beta(j)
  - \frac{B}2 \sum_i h_\beta^\alpha S_\alpha^\beta(i)
\label{21H3}
\end{equation}
with $h=((0,i,0),(-i,0,0),(0,0,0))$.

To generalize the model to O($N$), we introduce
$N$ Bose fields $b_\alpha$ subject to the constraints
\begin{eqnarray}
b_\alpha^\dagger b_\alpha & = & \frac{NS}{3} ,
\label{21c4a} \\
b_\alpha^\dagger b_\alpha^\dagger & = & 0 .
\label{21c4b}
\end{eqnarray}
The O($N$) spin matrices are
$S_\beta^\alpha = b_\alpha^\dagger b_\beta - b_\beta^\dagger b_\alpha$.
The second constraint again restricts the Hilbert space
by identifying, say, $b_N^\dagger b_N^\dagger|\psi\rangle$ with another
state. The O($N$) Hamiltonian is
\begin{equation}
H = -\frac{3J}{2N} \sum_{\langle ij\rangle} S_\beta^\alpha(i)
  S_\alpha^\beta(j)
  - \frac{B}2 \sum_i h_\beta^\alpha S_\alpha^\beta(i) ,
\label{21H4}
\end{equation}
where $h$ contains $N/3$ copies of the O(3) matrix
along the diagonal. $NS$ be an integer multiple of $3$.

The next steps are similar to the SU($N$) model. Going over to the
continuum and inserting bosons we get
\begin{eqnarray}
H & = & \int d^2r\, \bigg[ JS (\partial_j b_\alpha^\dagger)
    (\partial_j b_\alpha)
  - \frac{3J}{N} b_\alpha^\dagger(\partial_j b_\beta^\dagger)
    b_\beta(\partial_j b_\alpha) \nonumber \\
& & {}- \frac{B}{a^2} h_\beta^\alpha b_\beta^\dagger b_\alpha \bigg] .
\label{21H8}
\end{eqnarray}
In writing the partition function as a functional integral,
the first constraint (\ref{21c4a}) is implemented using a Lagrange
multiplier field $\lambda$. Two Lagrange multipliers $\mu_1$ and $\mu_2$
are introduced to enforce the two components of the second constraint
(\ref{21c4b}). They couple to the $b_\alpha$ fields in the form
$\mu_1\,\text{Re}\, b_\alpha b_\alpha
  + \mu_2\,\text{Im}\, b_\alpha b_\alpha
= \mu^\ast\,b_\alpha b_\alpha/2 + \mu\,b_\alpha^\ast b_\alpha^\ast/2$,
where we have introduced $\mu=\mu_1+i\mu_2$, which is somewhat
misleading, though, since both $\mu_1$ and $\mu_2$ have to be integrated
along the {\it imaginary\/} axis. The partition function reads
\begin{equation}
Z = \int D^2b_\alpha\,D\lambda\,D^2\mu \exp\!\left( -\frac1{\hbar}
  \int_0^{\hbar\beta}\!\!\! d\tau
  \int d^2r\, {\cal L}[b;\lambda,\mu] \right) ,
\end{equation}
where
\begin{eqnarray}
{\cal L} & = & \frac{\hbar}{a^2} b_\alpha^\ast \partial_0 b_\alpha
  + JS (\partial_j b_\alpha^\ast)(\partial_j b_\alpha)
  - \frac{3J}{N} b_\alpha^\ast(\partial_j b_\beta^\ast)
    b_\beta(\partial_j b_\alpha) \nonumber \\
& & {}- \frac{B}{a^2} h_\beta^\alpha b_\beta^\ast b_\alpha
  + \lambda b_\alpha^\ast b_\alpha - \frac{NS}{3} \lambda
  \nonumber \\
& & {}+ \frac12 \mu^\ast\,b_\alpha b_\alpha
  + \frac12 \mu\,b_\alpha^\ast b_\alpha^\ast .
\label{21L2}
\end{eqnarray}
The quartic term is decoupled using a Hubbard-Stra\-to\-no\-vich
transformation,
\begin{eqnarray}
{\cal L}' & = & \frac{\hbar}{a^2} b_\alpha^\ast \partial_0 b_\alpha
  + JS (\partial_j b_\alpha^\ast)(\partial_j b_\alpha)
  + 3NJ Q_jQ_j  \nonumber \\
& & {}+ 3iJ Q_j b_\alpha^\ast(\partial_j b_\alpha)
  - 3iJ Q_j (\partial_j b_\alpha^\ast)b_\alpha
  - \frac{B}{a^2} h_\beta^\alpha b_\beta^\ast b_\alpha  \nonumber \\
& & {}+ \lambda b_\alpha^\ast b_\alpha - \frac{NS}{3} \lambda
  + \frac12 \mu^\ast\,b_\alpha b_\alpha
  + \frac12 \mu\,b_\alpha^\ast b_\alpha^\ast ,
\label{21L4}
\end{eqnarray}
where $Q_j$ is real and a gauge field. As compared to SU($N$),
additional complications arise since under gauge changes $\mu$
transforms like a charge $2$ particle,
as discussed below. We choose the transverse gauge, $\partial_jQ_j=0$.

\subsection{Mean field theory}
\label{3.2}

Again, MF theory is exact for $N\to\infty$.
We make a static assumption for $\lambda$, ${\bf Q}$, and $\mu$.
We then express the fields $b_\alpha$ in terms of Fourier transforms.
The partition function reads
\begin{eqnarray}
Z_0 & = & \int D^2b_\alpha({\bf k},i\omega_n) \exp\!\bigg( -3{\cal N}
  N\beta J \overline{\bf Q}\cdot
  \overline{\bf Q} a^2 \nonumber \\
& & {}+ {\cal N} a^2\frac{NS}{3} \beta{\overline{\lambda}} -
    \int d^2k \sum_{i\omega_n} {\cal L}''_0[b] \bigg)
\end{eqnarray}
with
\begin{eqnarray}
{\cal L}''_0[b] & = & \beta a^2 \sum_\alpha \left(
  - i\hbar\omega_n + JS k^2a^2 - 6J \overline{\bf Q}\cdot{\bf k} a^2
  + a^2 {\overline{\lambda}} \right) b_\alpha^\ast({\bf k},i\omega_n)
    b_\alpha({\bf k},i\omega_n)  \nonumber \\
& & {}- \beta a^2 \sum_{\alpha\beta} B h_\beta^\alpha\,
    b_\beta^\ast({\bf k},i\omega_n)b_\alpha({\bf k},i\omega_n)
  + \beta a^2 \sum_\alpha \frac{a^2}2 {\overline{\mu}}^\ast\,
    b_\alpha(-{\bf k},-i\omega_n)b_\alpha({\bf k},i\omega_n)
  \nonumber \\
& & {}+
  \beta a^2 \sum_\alpha \frac{a^2}2 {\overline{\mu}}\:
    b_\alpha^\ast(-{\bf k},-i\omega_n)b_\alpha^\ast({\bf k},i\omega_n) ,
\label{22L3}
\end{eqnarray}
where sums are again written out.
To diagonalize the Zeeman term we substitute new fields,
\begin{equation}
c_{3n+1} = \frac1{\sqrt{2}} (b_{3n+1}+ib_{3n+2}) ,\quad
c_{3n+2} = \frac1{\sqrt{2}} (b_{3n+1}-ib_{3n+2}) ,\quad
c_{3n+3} = b_{3n+3} .
\label{22dc1a}
\end{equation}
After shifting the integration variable ${\bf k}$ to
${\bf k}+3\overline{\bf Q}/S$ we get
\begin{eqnarray}
{\cal L}''_0[c] & = & \beta a^2 \sum_\alpha \left(
  - i\hbar\omega_n + JS k^2a^2 - B \hat{h}_\alpha^\alpha
  + a^2 {\overline{\lambda}} 
  - \frac{9J}{S} \overline{\bf Q}\cdot\overline{\bf Q} a^2 \right)
    c_\alpha^\ast({\bf k},i\omega_n)c_\alpha({\bf k},i\omega_n)
  \nonumber \\
& & {}+ \beta a^2 \sum_\alpha \frac{a^2}2 {\overline{\mu}}^\ast\,
    c_{\overline{\alpha}}(-{\bf k},-i\omega_n)
    c_\alpha({\bf k},i\omega_n)
  + \beta a^2 \sum_\alpha \frac{a^2}2 {\overline{\mu}}\:
    c_{\overline{\alpha}}^\ast(-{\bf k},-i\omega_n)
    c_\alpha^\ast({\bf k},i\omega_n) ,
\label{22L4}
\end{eqnarray}
where $\hat{h}$ is diagonal with the diagonal elements
$-1,1,0,-1,1,0,\ldots$ and
\begin{equation}
\overline{\alpha} = \left\{\begin{array}{ll}
  3n+2 & \mbox{for } \alpha = 3n+1 \\
  3n+1 & \mbox{for } \alpha = 3n+2 \\
  3n+3 & \mbox{for } \alpha = 3n+3.
  \end{array} \right.
\end{equation}
The partition function depends on ${\overline{\lambda}}$ and
$\overline{\bf Q}$ only through
${\overline{\Lambda}}
  \equiv a^2 \beta{\overline{\lambda}} - (9\beta J/S)
  \overline{\bf Q}\cdot\overline{\bf Q} a^2$.
To get rid of the terms mixing $({\bf k},i\omega_n)$ with
$(-{\bf k},-i\omega_n)$ we note that $c_\alpha({\bf k},i\omega_n)$ is
even in $\omega_n$ and introduce new fields,
\begin{equation}
d_\alpha({\bf k},i\omega_n) =
  \frac1{\sqrt{2}} [c_\alpha({\bf k},i\omega_n)
  - i c_\alpha(-{\bf k},i\omega_n)] .
\label{22d5}
\end{equation}
Then we have
\begin{eqnarray}
{\cal L}''_0[d] & = & \beta a^2 \sum_\alpha \left(
  - i\hbar\omega_n + JS k^2a^2 - B \hat{h}_\alpha^\alpha
  + \frac{{\overline{\Lambda}}}{\beta} \right)
    d_\alpha^\ast({\bf k},i\omega_n)d_\alpha({\bf k},i\omega_n)
  \nonumber \\
& & {}+ \beta a^2 \sum_\alpha \frac{ia^2}2 {\overline{\mu}}^\ast\,
    d_{\overline{\alpha}}({\bf k},i\omega_n)d_\alpha({\bf k},i\omega_n)
  - \beta a^2 \sum_\alpha \frac{ia^2}2 {\overline{\mu}}\:
    d_{\overline{\alpha}}^\ast({\bf k},i\omega_n)
    d_\alpha^\ast({\bf k},i\omega_n) .
\label{22L5}
\end{eqnarray}
The fields $d_\alpha$ are now integrated out. We define
${\tilde J} \equiv \beta J$ and ${\tilde B} \equiv \beta B$
(note different definition of ${\tilde B}$).
By integrating over the $d_\alpha$, putting the product into the
exponential, and evaluating the flavor sum we obtain
\begin{eqnarray}
Z_0 & \propto & \exp\!\Bigg[ {\cal N} \frac{NS}{3}{\overline{\Lambda}}
  - \frac{{\cal N} a^2}{4\pi^2}\,\frac{N}{3} \int d^2k
  \sum_{i\omega_n} \bigg( \frac12 \ln \left[
    (-i\beta\hbar\omega_n + {\tilde J} S k^2a^2
    + {\overline{\Lambda}})^2 - a^4\beta^2{\overline{\mu}}^\ast
    {\overline{\mu}} \right]  \nonumber \\
& & {}+ \ln \left[ (-i\beta\hbar\omega_n + {\tilde J} S k^2a^2
  + {\overline{\Lambda}})^2
  - a^4 \beta^2 {\overline{\mu}}^\ast{\overline{\mu}}
  - {\tilde B}^2 \right] \bigg) \Bigg] .
\label{22Z7}
\end{eqnarray}
The MF values ${\overline{\Lambda}}$ and ${\overline{\mu}}$ are
determined by the saddle-point equations
\begin{equation}
\frac{\partial\ln Z_0}{\partial{\overline{\Lambda}}} = 0  ,\qquad
\frac{\partial\ln Z_0}{\partial{\overline{\mu}}} = 0  ,\qquad
\frac{\partial\ln Z_0}{\partial{\overline{\mu}}^\ast} = 0 .
\end{equation}
The last two are equivalent. They yield
\begin{eqnarray}
0 & = & a^4\beta^2{\overline{\mu}} \int d^2k
  \sum_{i\omega_n} \bigg( \frac12\,
    \frac1{(-i\beta\hbar\omega_n + {\tilde J} S k^2a^2
  + {\overline{\Lambda}})^2 - a^4\beta^2{\overline{\mu}}^\ast
    {\overline{\mu}}} \nonumber \\
& & {}+ \frac1{(-i\beta\hbar\omega_n + {\tilde J} S k^2a^2
  + {\overline{\Lambda}})^2
  - a^4 \beta^2 {\overline{\mu}}^\ast{\overline{\mu}} - {\tilde B}^2}
  \bigg) .
\end{eqnarray}
One solution is ${\overline{\mu}}=0$. For ${\overline{\mu}}\neq0$ we
get
\begin{eqnarray}
1 & = & \left[1-2e^{-{\overline{\Lambda}}}\cosh(a^2\beta|{\overline{\mu}}|)
  + e^{-2{\overline{\Lambda}}}\right] \nonumber \\
& & \times \left[1-2e^{-{\overline{\Lambda}}}
    \cosh\sqrt{a^4\beta^2|{\overline{\mu}}|^2+{\tilde B}^2}
  + e^{-2{\overline{\Lambda}}}\right]^2 ,
\end{eqnarray}
which has solutions for $|{\overline{\mu}}|>0$, corresponding to broken
gauge symmetry. The saddle-point equation for ${\overline{\Lambda}}$
yields
\begin{eqnarray}
\lefteqn{
e^{-8\pi{\tilde J} S^2} = \left[1-2e^{-{\overline{\Lambda}}}
  \cosh(a^2\beta|{\overline{\mu}}|)+e^{-2{\overline{\Lambda}}}\right]
  } \nonumber \\
& & \qquad\times \left[1-2e^{-{\overline{\Lambda}}}
    \cosh\sqrt{a^4\beta^2|{\overline{\mu}}|^2+{\tilde B}^2}
  + e^{-2{\overline{\Lambda}}}\right]^2
\end{eqnarray}
so that there are no simultaneous solutions with ${\overline{\mu}}\neq 0$.
Note that {\it if\/} we had ${\overline{\mu}}\neq 0$,
a term like $|(\partial_j - 6Q_j/S)\mu|^2$ would appear in the gauge
invariant Lagrangian, which would make the gauge field ${\bf Q}$
massive\cite{SuSpriv} (this is the Anderson-Higgs mechanism).
In our case, however, it is massless at the saddle point.

The partition function is now
\begin{eqnarray}
Z_0 & \propto & \exp\!\Bigg[ {\cal N} \frac{NS}{3}{\overline{\Lambda}}
  - \frac{{\cal N} a^2}{4\pi^2} \int d^2k
  \sum_{i\omega_n} \sum_\alpha \nonumber \\
& & \times \ln \left( -i\beta\hbar\omega_n
    + {\tilde J} S k^2a^2
    - {\tilde B} \hat{h}_\alpha^\alpha + {\overline{\Lambda}} \right)
  \Bigg] .
\end{eqnarray}
The MF equation for ${\overline{\Lambda}}$ becomes\cite{RS}
\begin{eqnarray}
S & = & -\frac1{4\pi{\tilde J} S} \Big[
  \ln(1-e^{-{\overline{\Lambda}}+{\tilde B}})
  + \ln(1-e^{-{\overline{\Lambda}}}) \nonumber \\
& & {}+ \ln(1-e^{-{\overline{\Lambda}}-{\tilde B}}) \Big]
\label{22s6}
\end{eqnarray}
and the MF magnetization is\cite{RS}
\begin{eqnarray}
M_0 & = & \frac{3}{{\cal N} N\beta} \frac{d}{dB} \ln Z_0
  \nonumber \\
& = & -\frac1{4\pi{\tilde J} S} \left[
  \ln(1-e^{-{\overline{\Lambda}}+{\tilde B}})
  - \ln(1-e^{-{\overline{\Lambda}}-{\tilde B}}) \right] ,
\label{22M6}
\end{eqnarray}
which exhibits the same universality as the SU($N$) result.
At low temperatures,
\begin{equation}
M_0 - S \cong \frac1{4\pi{\tilde J} S}\,\ln(1-e^{-\beta B})
  + \frac1{2\pi{\tilde J} S}\,\ln(1-e^{-2\beta B})
\end{equation}
up to exponentially small corrections to the magnetic field. Thus,
although the leading term is the same as the non-interacting
magnon approximation, Eq.~(\ref{12M8}),
the second term is different. Thus we expect the correct behavior
at the lowest temperatures but deviations already for $T\sim 2B$.

\subsection{$1/N$ corrections}
\label{3.3}

The method used to calculate the $1/N$ corrections
is similar to the SU($N$) case. However, the second constraint
(\ref{21c4b}) introduces additional problems. The magnetization is now
\begin{equation}
M = \frac3{N} \sum_{\alpha\beta} h_\beta^\alpha \langle
  b_\alpha^\dagger b_\beta\rangle
= \frac3{N} \sum_\alpha \hat{h}_\alpha^\alpha \langle
  c_\alpha^\dagger c_\alpha\rangle ,
\end{equation}
using the definition of $c_\alpha$ in Eq.~(\ref{22dc1a}).
Fourier transforming the $c_\alpha$ we find
$\langle c_\alpha^\dagger c_\alpha\rangle
 = \langle d_\alpha^\dagger d_\alpha\rangle$ (no summation implied) and
\begin{equation}
M = \frac3{N} \sum_\alpha \hat{h}_\alpha^\alpha \langle
  d_\alpha^\dagger d_\alpha\rangle .
\label{23M1}
\end{equation}
In the following we use the representation in terms of fields
$d_\alpha$, Eq.~(\ref{22d5}). The fluctuations are written as
\begin{eqnarray}
\lambda({\bf r},\tau) & = & {\overline{\lambda}}
  + i\Delta\lambda({\bf r},\tau) ,  \\
\mu_1({\bf r},\tau) & = & 0 + i\Delta\mu_1({\bf r},\tau) ,
\label{23f2}  \\
\mu_2({\bf r},\tau) & = & 0 + i\Delta\mu_2({\bf r},\tau) ,
\label{23f3}  \\
Q_j({\bf r},\tau) & = & 0 + \Delta Q_j({\bf r},\tau) ,
\end{eqnarray}
where $\Delta\lambda$, $\Delta Q_j$, $\Delta\mu_1$, and $\Delta\mu_2$
are all real. For convenience we use a complex
$\Delta\mu=\Delta\mu_1+i\Delta\mu_2$ so that
$\mu({\bf r},\tau) = i\Delta\mu({\bf r},\tau)$ and
$\mu^\ast({\bf r},\tau) = -i\Delta\mu^\ast({\bf r},\tau)$.
The SU($N$) methods of Ref.~\onlinecite{Auerbach} can be adapted
to the O($N$) model; we write
\begin{equation}
Z = \int D\Delta\lambda\,D^2\Delta\mu\,D\Delta Q_j \exp(-N{\cal S})
\end{equation}
and expand the action ${\cal S}$ as in Eq.~(\ref{13A1}) for SU($N$),
where $r_\ell$
can also stand for $\Delta\mu$ or $\Delta\mu^\ast$. We can also write
${\cal S}={\cal S}_0+{\cal S}_{\text{dir}}+{\cal S}_{\text{loop}}$ with
\begin{eqnarray}
{\cal S}_0 & = & \frac1{N}\,\text{Tr}\ln G_0^{-1} ,  \\
{\cal S}_{\text{dir}} & = & \frac1{N\hbar} \int_0^{\hbar\beta}\!\! d\tau
  \int d^2r\, \left(3NJ {\bf Q}\cdot{\bf Q}
  - \frac{NS}{3} \lambda\right) ,\!\!  \\
{\cal S}_{\text{loop}} & = & \frac1{N}\,\text{Tr}\ln\!\left(
  1 + G_0 \sum_\ell \upsilon_\ell r_\ell \right) .
\end{eqnarray}
The (normal) MF Green function can be read off from Eq.~(\ref{22L5}),
\begin{equation}
G_0^\alpha({\bf k},i\omega_n) =
  (-i\beta\hbar\omega_n+{\tilde J} S k^2a^2
  -{\tilde B} \hat{h}_\alpha^\alpha+{\overline{\Lambda}})^{-1} .
\end{equation}
For ${\overline{\mu}}=0$ no anomalous Green function exists
since there is no $d_\alpha^\ast d_\alpha^\ast$ term.

The constant part of ${\cal S}_{\text{dir}}$ together with ${\cal S}_0$
again gives the MF action ${\cal S}^{(0)}$. The first-order terms
cancel. The vertex factors can be found in analogy to SU($N$),
\begin{eqnarray}
& & \upsilon_{\Delta\lambda} = \frac{2\pi}{{\cal N}} i a^2 ,\quad
  \upsilon_{\Delta\mu} = \frac{2\pi}{{\cal N}} \frac{i}{2} a^2 ,\quad
  \upsilon_{\Delta\mu^\ast} = \frac{2\pi}{{\cal N}} \frac{i}{2} a^2 ,
  \nonumber \\
& & \upsilon_{\Delta Q_j} = \frac{2\pi}{{\cal N}} (-6 J)\, a^2 k_j .
\end{eqnarray}
The diagrammatics are similar to the SU($N$) case. For $\Delta\lambda$
and $\Delta{\bf Q}$ fluctuations the only differences are: (i) The
$\Delta Q_j$ vertices contain an
additional factor of $3$ each, giving $9$ in ${\cal S}^{(2)}$,
(ii) the direct $\Delta Q_j$ propagator from ${\cal S}_{\text{dir}}$
contains an additional factor of $3$, (iii) $h_\alpha^\alpha$ is
replaced by $\hat{h}_\alpha^\alpha$, (iv) now
${\tilde B}\equiv \beta B$,
and (v) ${\overline{\Lambda}}$ is given by Eq.~(\ref{22s6}).

In particular, we find
${\cal S}_{\Delta Q_2(0,0),\Delta Q_2(0,0)}^{(2)} = 0$
as for the SU($N$) model. Thus gauge fluctuations are massless for
O($N$) as well as for SU($N$).

\begin{figure}[htb]
\centerline{
\epsfig{file=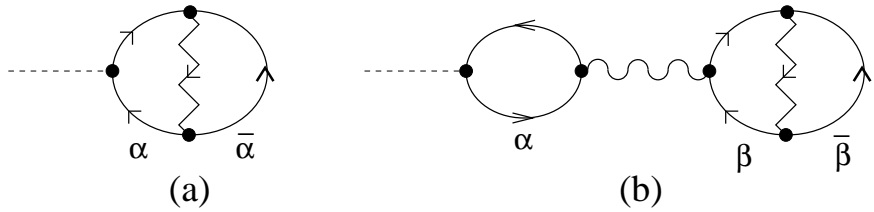,width=4.5in}}
\caption{Diagrams in the $1/N$ magnetization containing
$\Delta\mu$ fluctuations.}
\label{fig230.5}
\end{figure}

The contribution from $\Delta\mu$ requires some thought. From
Eqs.~(\ref{21L4}) and (\ref{22L5}) we see that $\Delta\mu$ couples to
two ``creation operators'' $d_\alpha^\ast d_{\overline{\alpha}}^\ast$,
whereas $\Delta\mu^\ast$ couples to $d_\alpha d_{\overline{\alpha}}$.
Consequently,
the boson loop in ${\cal S}^{(2)}$ can only contain one $\Delta\mu$
vertex and one $\Delta\mu^\ast$ vertex or neither of them. Thus
${\cal S}^{(2)}$ and the RPA propagator do not mix
$\Delta\mu$ with other fluctuations. Consequently, the only
contributions to $\langle d_\alpha^\dagger d_\alpha\rangle$ at the
$1/N$ level correspond to the diagrams in Fig.~\ref{fig230.5}, where
the zig-zag line denotes the $\Delta\mu$ RPA propagator. Note the
directions of the boson lines.

We derive ${\cal S}_{\Delta\mu^\ast,\Delta\mu}^{(2)}$,
${\cal S}_{\Delta\lambda(0,0),\Delta\mu^\ast,\Delta\mu}^{(3)}$, and
${\cal S}_{j_\alpha;\Delta\mu^\ast,\Delta\mu}^{(2+1)}$
in App.~\ref{appB}. We can then integrate out the fluctuations.
$\Delta\mu$ is a complex field so that the contraction of a pair yields
\begin{equation}
\frac1{Z} \int D^2z_\ell\, z_{\ell_1}^\ast z_{\ell_2} \exp\!
  \left(-\frac{N}{2}
  {\cal S}_{\ell_1'\ell_2'}^{(2)} z_{\ell_1'}^\ast z_{\ell_2'}\right)
  = \frac2{N} ({\cal S}^{(2)})^{-1}_{\ell_1\ell_2} .
\end{equation}
Consequently, the diagrams of Figs.~\ref{fig132.5}(a) and
\ref{fig230.5}(a) added together, and Figs.~\ref{fig132.5}(b) plus
\ref{fig230.5}(b), respectively, are
\begin{eqnarray}
{\cal D}_1^{(0)} & = & \frac1{2{\cal N}} \sum_{{\bf q},i\nu_n} \left(
  \frac{\sigma_0^{\alpha\prime}}{\sum_\beta \sigma_0^\beta} +
  \frac{\sigma_\perp^{\alpha\prime}}{{\cal N} N/6{\tilde J}+\sum_\beta
  \sigma_\perp^\beta} +
 2\,\frac{\sigma_\star^{\alpha\prime}}{\sum_\beta \sigma_\star^\beta}
  \right) , \\
{\cal D}_2^{(0)} & = & -\frac{3}{2{\cal N} N}\,
  \frac{n_B({\overline{\Lambda}}-{\tilde B} \hat{h}_\alpha^\alpha)}
  {n_B({\overline{\Lambda}}-{\tilde B})+n_B({\overline{\Lambda}})
    +n_B({\overline{\Lambda}}+{\tilde B})}  \nonumber \\
& & \times \sum_{{\bf q},i\nu_n} \left(
  \frac{\sum_\beta \sigma_0^{\beta\prime}}{\sum_\beta \sigma_0^\beta} +
  \frac{\sum_\beta \sigma_\perp^{\beta\prime}}
    {{\cal N} N/6{\tilde J}+\sum_\beta \sigma_\perp^\beta} +
 2\,\frac{\sum_\beta \sigma_\star^{\beta\prime}}{\sum_\beta
   \sigma_\star^\beta} \right) ,
\end{eqnarray}
where
\begin{eqnarray}
\sigma_\star^\alpha & \equiv & \sum_{\bf k}
  \frac{1+n_B(\epsilon_{{\bf k}+{\bf q}/2}^\alpha)
  +n_B(\epsilon_{{\bf k}-{\bf q}/2}^\alpha)}
    {-i\beta\hbar\nu_n+2{\tilde J} S\,k^2a^2
    +{\tilde J} S\,q^2a^2/2+2{\overline{\Lambda}}} ,
\label{23d10} \\
\sigma_\star^{\alpha\prime} & \equiv & \sum_{\bf k} \Bigg(
  \frac{n_B^{(1)}(\epsilon_{{\bf k}+{\bf q}/2}^\alpha)
  +n_B^{(1)}(\epsilon_{{\bf k}-{\bf q}/2}^\alpha)}
    {-i\beta\hbar\nu_n+2{\tilde J} S\,k^2a^2
    +{\tilde J} S\,q^2a^2/2+2{\overline{\Lambda}}}
  - 2\frac{1+n_B(\epsilon_{{\bf k}+{\bf q}/2}^\alpha)
    +n_B(\epsilon_{{\bf k}-{\bf q}/2}^\alpha)}
    {(-i\beta\hbar\nu_n+2{\tilde J} S\,k^2a^2
    +{\tilde J} S\,q^2a^2/2+2{\overline{\Lambda}})^2} \Bigg) ,
\label{23d11}
\end{eqnarray}
the other symbols are identical to the SU($N$) case if $h_\alpha^\alpha$
is replaced by $\hat{h}_\alpha^\alpha$.

With Eq.~(\ref{23M1}) the $1/N$ contribution to the magnetization reads
\begin{equation}
M = M_0 - \frac1{N}\,\frac{3a^2}{8\pi^2} \int d^2q \sum_{i\nu_n} \left(
  \frac{\sum_\alpha (c_1-\hat{h}_\alpha^\alpha)
    \sigma_0^{\alpha\prime}}{\sum_\beta \sigma_0^\beta} +
  \frac{\sum_\alpha (c_1-\hat{h}_\alpha^\alpha)
    \sigma_\perp^{\alpha\prime}}
  {{\cal N} N/6{\tilde J}+\sum_\beta \sigma_\perp^\beta} +
  2\,\frac{\sum_\alpha (c_1-\hat{h}_\alpha^\alpha)
    \sigma_\star^{\alpha\prime}}
  {\sum_\beta \sigma_\star^\beta} \right)
\label{23M8}
\end{equation}
with 
\begin{equation}
c_1 \equiv \frac{n_B({\overline{\Lambda}}-{\tilde B})
  - n_B({\overline{\Lambda}}+{\tilde B})}
  {n_B({\overline{\Lambda}}-{\tilde B}) + n_B({\overline{\Lambda}})
  + n_B({\overline{\Lambda}}+{\tilde B})} .
\end{equation}
Evaluation of the $\Delta\lambda$ and $\Delta Q_2$ contributions
is analogous to the SU($N$) case. In particular, naive summation over
$i\nu_n$ results in a strong divergence. The constant term in the
integrand for large momenta is $3(c_1 - M_0/S)$. Numerical calculations
confirm this result. Again, the spurious divergence is removed by
taking operator ordering into account.

We now turn to the $\Delta\mu$ contribution. From Eq.~(\ref{23d10}) we
see that $\sigma_\star^\alpha$ diverges logarithmically at large
momentum ${\bf k}$ because of the summand $1$ in the numerator.
However, the $\Delta\mu$ contribution to the magnetization is finite.
To see this, we use a finite cutoff $K$ and let $K\to\infty$ in the
result. $\sigma_\star^\alpha$ is dominated by
\begin{eqnarray}
\sigma_\star^\alpha & \cong & \frac{{\cal N} a^2}{4\pi^2}
  \int_{k\le K}\!\! d^2k\, \frac1
  {-i\beta\hbar\nu_n+2{\tilde J} S\,k^2a^2
    +{\tilde J} S\,q^2a^2/2+2{\overline{\Lambda}}}  \nonumber \\
& = & \frac{{\cal N}}{8\pi{\tilde J} S} \left[
  \ln(-i\beta\hbar\nu_n+2{\tilde J} S\,K^2a^2
    +{\tilde J} S\,q^2a^2/2+2{\overline{\Lambda}})
  - \ln(-i\beta\hbar\nu_n+{\tilde J} S\,q^2a^2/2+2{\overline{\Lambda}})
  \right] .
\label{23x12a}
\end{eqnarray}
For $\sigma_\star^{\alpha\prime}$ the corresponding contribution is
\begin{equation}
\sigma_\star^{\alpha\prime} \cong \frac{-2{\cal N}}{8\pi{\tilde J} S}
  \left( \frac1{-i\beta\hbar\nu_n+2{\tilde J} S\,K^2a^2
    +{\tilde J} S\,q^2a^2/2+2{\overline{\Lambda}}}
  - \frac1{-i\beta\hbar\nu_n+{\tilde J} S\,q^2a^2/2
    +2{\overline{\Lambda}}} \right) .
\end{equation}
Note that these two expressions do not depend on $\alpha$.
The frequency sum over the $\Delta\mu$ contribution is
\begin{eqnarray}
\lefteqn{
\sum_{i\nu_n} \Delta M_\star({\bf q},i\nu_n)
\equiv \sum_{i\nu_n} \frac{\sum_\alpha (c_1-\hat{h}_\alpha^\alpha)
  \sigma_\star^{\alpha\prime}} {\sum_\alpha \sigma_\star^\alpha}
  } \nonumber \\
& & = -2c_1 \sum_{i\nu_n} \frac{\displaystyle
  \frac1{-i\beta\hbar\nu_n+\epsilon_K}
    - \frac1{-i\beta\hbar\nu_n+\epsilon_0}}
  {\ln(-i\beta\hbar\nu_n+\epsilon_K)
    - \ln(-i\beta\hbar\nu_n+\epsilon_0)} .
\end{eqnarray}
where
$\epsilon_k\equiv
  2{\tilde J} S\,k^2a^2+{\tilde J} S\,q^2a^2/2+2{\overline{\Lambda}}$.
Since this contribution is proportional to $c_1$ it
comes only from the diagram Fig.~\ref{fig230.5}(b).

The sum over $i\nu_n$ can be evaluated by contour integration. As noted
in App.~\ref{appB}, splitting the time to enforce correct operator
ordering results in an overall factor of
$\exp(i\nu_n\eta)$, which removes any ambiguity in the $i\nu_n$ sum.
In the complex $\nu$ plane, $\Delta M_\star$ has a branch cut along the
real axis between the points $\epsilon_0/\hbar\beta$ and
$\epsilon_K/\hbar\beta$ and two poles on top of the branch points.
The contour integral contains three terms: Two
from integrating around the branch points in small semicircles and one
from integrating the spectral function of $\Delta M_\star$ along the
branch cut. The two semicircles contribute
$-2c_1 [ n_B(\epsilon_0) + n_B(\epsilon_K) ]$.
For $K\to\infty$ this expression
becomes $-2c_1 n_B(\epsilon_0)$. The spectral function is
\begin{equation}
{\cal A}_\star = -\frac{4\pi c_1}{\hbar\beta}\, \frac{\displaystyle
  P\frac1{\epsilon_K/\hbar\beta-\nu}
    + P\frac1{\nu-\epsilon_0/\hbar\beta}}
  {\left[\ln(\epsilon_K/\hbar\beta-\nu)
    - \ln(\nu-\epsilon_0/\hbar\beta)\right]^2 + \pi^2}
\end{equation}
and the integral over it can be shown to vanish for $K\to\infty$. Thus
\begin{equation}
\sum_{i\nu_n} \Delta M_\star
  = -2c_1 n_B({\tilde J} S\,q^2a^2/2+2{\overline{\Lambda}})
\end{equation}
for $K\to\infty$.  With Eq.~(\ref{23M8})
the full contribution from $\Delta\mu$ to the magnetization is
\begin{eqnarray}
\lefteqn{
-\frac1{N}\,\frac{3a^2}{8\pi^2} \int d^2q \sum_{i\nu_n} 2 \Delta M_\star
  = -\frac1{N}\,\frac{3}{\pi{\tilde J} S}\,
  \ln\!(1-e^{-2{\overline{\Lambda}}}) } \nonumber \\
& & \qquad\times \frac{n_B({\overline{\Lambda}}-{\tilde B})
    - n_B({\overline{\Lambda}}+{\tilde B})}
  {n_B({\overline{\Lambda}}-{\tilde B})+n_B({\overline{\Lambda}})
    + n_B({\overline{\Lambda}}+{\tilde B})} .
\end{eqnarray}
A few remarks are in order: (i) This contribution {\it increases\/} the
magnetization, whereas fluctuations in $\lambda$ and ${\bf Q}$ decrease
it. The physical explanation is that the MF approximation, which
enforces the second constraint $b_\alpha^\dagger b_\alpha^\dagger=0$
only on average, {\it under\/}estimates
the magnetization because it contains contributions from spurious
multiplets of lower total spin.
(ii) The $\Delta\mu$ contribution has a typical energy scale of
$2{\overline{\Lambda}}$ since excitations of energy
$2{\overline{\Lambda}}$ (and higher) are removed by the second
constraint (\ref{21c4b}).
(iii) The ${\bf q}$ integral over $\Delta M_\star$ is well-behaved for
large ${\bf q}$ so that a cutoff, which is necessary
for the $\Delta\lambda$ and $\Delta{\bf Q}$ contributions,
does not change the result appreciably but would complicate
the calculations.

\begin{figure}[tb]
\centerline{
\epsfig{file=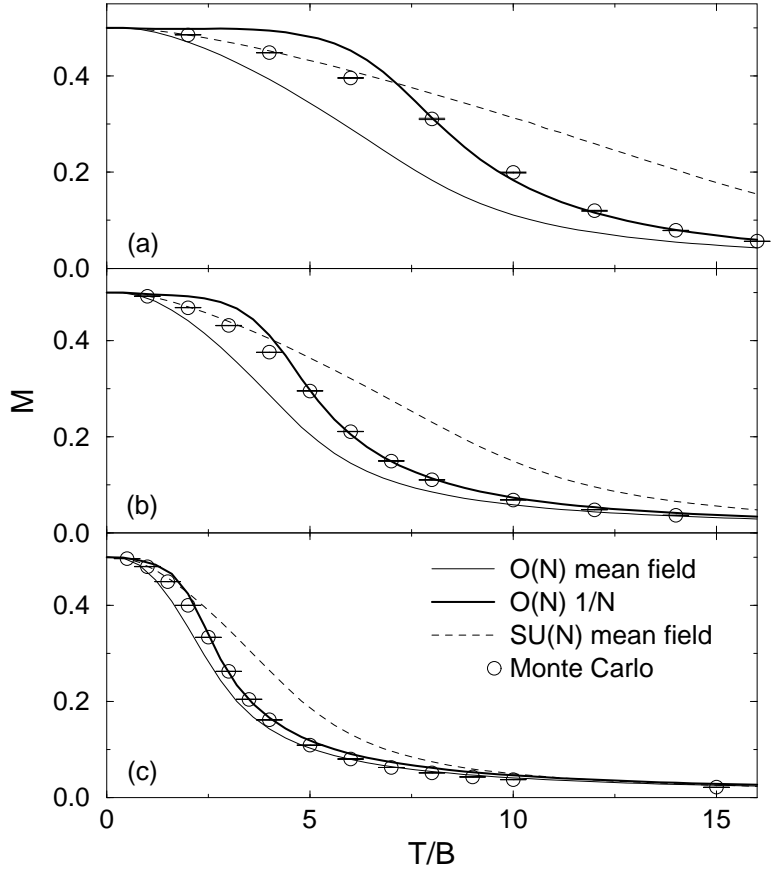,width=4in}}
\caption{O($N$) magnetization for magnetic fields (a) $B/J=0.05$,
(b) $B/J=0.1$, and (c) $B/J=0.25$. The thin solid line is the MF
magnetization, the thick solid curve includes $1/N$ corrections, the
circles with error bars are quantum Monte Carlo results for a
$32\times32$ lattice,\protect\cite{letter,PH} and the dashed curve
shows the SU($N$) MF magnetization for comparison.}
\label{fig232}
\end{figure}

Figure \ref{fig232} shows the magnetization
for $S=1/2$ and $B/J=0.05, 0.1, 0.25$ as a function of $T/B$.
The $\Delta\mu$ fluctuations win over the other contributions;
the magnetization is larger than the MF result.
We see that the O($N$) $1/N$ expansion gives much better results than
the SU($N$) model except at low temperatures. At small $T/B$ the
magnetization seems to be unphysically large, especially for smaller
fields. At moderate temperatures the O($N$) $1/N$ magnetization
is better than both MF results. At high temperatures, the Monte Carlo
data consistently
fall slightly below the O($N$) $1/N$ and, for $B/J=0.25$, even below
the O($N$) MF results.

Recall that the SU($N$) MF approximation works well at low temperatures
because it coincides with the non-interacting magnon approximation for
the Heisenberg model, whereas the O($N$) MF magnetization does not. The
$1/N$ corrections for the O($N$) model (predominantly from $\Delta\mu$
at low $T$) are large and
in fact overcompensate for the error made at the MF level.

There is a distinct crossover to the moderate $T$ regime, where SU($N$)
MF becomes too large, SU($N$) $1/N$ becomes quite wrong, and O($N$)
$1/N$ is rather good. In fact it is surprisingly good considering that
$1/N$ is not really small.
It is not fully clear why the O($N$) model works better than
the SU($N$) model at moderate and high temperatures.
The reason may lie in the different behavior of gauge ($\Delta{\bf Q}$)
fluctuations in the two models.\cite{SuSpriv} In the SU($N$) model
they are massless in general, whereas for O($N$) they are
massless only because the MF value of $\mu$ happens
to vanish for this particular system. In both cases the zero mass
leads to an overestimate of fluctuations at the $1/N$ level. However,
in the O($N$) model fluctuations in $\mu$ are available to
compensate for this, thereby partly restoring the effect a massive
gauge field would have.

One might think that O($N$) should be worse since the O($N$) model for
$N>3$ does not have skyrmions, whereas the SU($N$) model has them for
all $N$.\cite{Raja} However, the $1/N$ expansion does not contain
these non-perturbative effects anyway.
On the other hand, they are, in principle, captured by the Monte Carlo
simulations.\cite{PH}

\begin{figure}[htb]
\centerline{
\epsfig{file=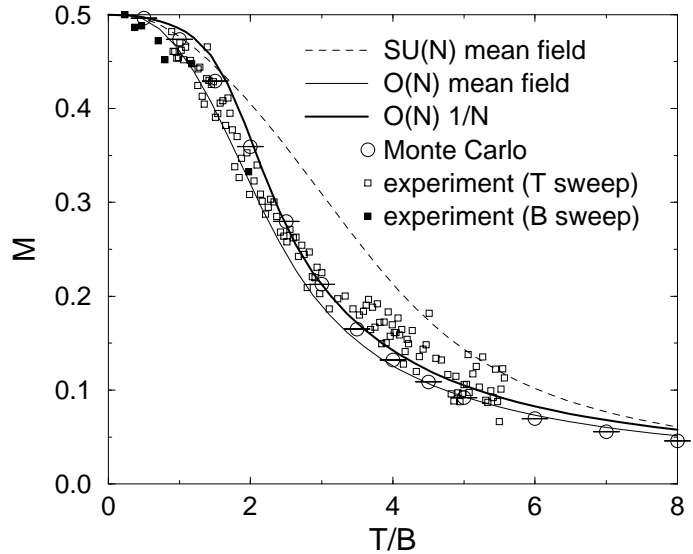,width=4in}}
\caption{Comparison of SU($N$) and O($N$) results, as well as quantum
Monte Carlo results for a $16\times16$ lattice, with experimental
data from Manfra {\it et al}.\protect\cite{Manfra}. The open squares
were obtained sweeping the temperature at fixed field and the filled
squares by sweeping the field at fixed temperature.}
\label{Mfig}
\end{figure}

The deviations between O($N$) results and Monte Carlo data at high
temperatures and $B/J=0.25$ or larger (see Fig.~\ref{Mfig}) are probably
due to thermally created skyrmions or to the fact that
the simulations are done on a lattice, whereas the $1/N$ calculations
use a continuum approximation. The dispersion of the former is a cosine
band if bosonized, whereas the latter has parabolic dispersion. Both
effects should become important for temperatures $T\ge J$ since both
the band width and the typical skyrmion energy are of the order of
$J$. Indeed, the deviations
start at $T\sim J$. (In the same region higher order gradient terms not
included in the Heisenberg model should become important.)

We have also investigated the universal dependence on ${\tilde J} S^2$
and ${\tilde B}$. Whereas the MF results exhibit this universality,
it is violated by small
logarithmic corrections at the $1/N$ level for both models,
as expected.\cite{RS}

Our results can be compared with the microscopic approach of
Kasner and MacDonald,\cite{KM} which includes spin-wave
corrections to the electronic self-energy. This approach is
microscopically better justified than the Heisenberg model. However,
the magnetization from Ref.~\onlinecite{KM} is consistently too large and
even MF SU($N$) and O($N$) results agree better with Monte Carlo
data.

Comparison to NMR experiments by Barrett {\it et al}.\cite{Barrett}
shows a number of discrepancies. At low temperature, the experimental
data look flat, whereas at high $T$ they drop well below the
theoretical results.
These discrepancies are mainly due to the scaling of the data, which
is done by setting the measured magnetization, which is reduced
by disorder, to $S$ in the limit $T\to0$.


Recent magnetoabsorption measurements
by Man\-fra {\it et al}.\cite{Manfra} show better
agreement with our results. In Fig.~\ref{Mfig} we compare data of
Ref.~\onlinecite{Manfra} with
SU($N$), O($N$), and quantum Monte Carlo results. In the calculations
we have used the exchange constant corrected for finite width of the
quantum well, which yields $B/J\approx 0.32$. The Monte Carlo data fall
below the O($N$) $1/N$
results above $T\sim J$, as discussed above. The experimental data
agree quite well with SU($N$) theory at low temperatures and with
O($N$) $1/N$ (and Monte Carlo) results at moderate temperatures, as
expected. At higher $T$ the experimental
data show more noise but lie mostly above the O($N$) $1/N$ curve.
This discrepancy for $T>J$ is probably
due to neglected higher gradient terms in Eq.~(\ref{11H1}).
The experimental system is a continuous itinerant magnet, which
probably explains the deviations from Monte Carlo lattice simulations.

\section{Summary and Conclusions}

We have calculated $1/N$ corrections to large $N$ Schwin\-ger boson
mean field theories for the two-di\-men\-sio\-nal ferromagnetic Heisenberg
model, meant to describe a quantum Hall system at filling factor
$\nu=1$. Normal ordering of operators has to be carefully taken into
account to obtain the corrections. Using a O($N$) model, we find
reasonable agreement of the $1/N$ corrected
magnetization with both quantum Monte Carlo simulations\cite{letter,PH}
and experiments\cite{Manfra} at moderate and higher temperatures.
At low temperatures, the SU($N$) model works better
since it reproduces the correct low-energy physics.
However, the SU($N$) model does not describe the data anywhere else,
confirming Auerbach's remark that large $N$ methods
are ``either surprisingly successful or completely
wrong.''\cite{Auerbach} Effects of thermally created skyrmions,
which are not included in our approach, are small. Away from filling
factor $\nu=1$, skyrmions are present in the ground state and should
be important. The natural next step leading on from this work would
be to incorporate these skyrmions. In addition higher derivative
terms due to the long range Coulomb interaction should be
investigated.

\acknowledgements

We wish to thank S. Sachdev, A.H. MacDonald, and A.W. Sandvik
for valuable discussions and B.B. Goldberg and M.J. Manfra
for helpful remarks and sharing their data.
This work has been supported by NSF DMR-9714055 and NSF CDA-9601632.
C.T. acknowledges support by the Deutsche Forschungsgemeinschaft and
P.H. by the Ella och Georg Ehrnrooths stiftelse.

\appendix


\section{Calculation of diagrams for SU($N$)}
\label{appA}

Here, we derive explicit expressions for ${\cal S}^{(2)}$,
${\cal S}^{(3)}$,
${\cal S}^{(1+1)}$, and ${\cal S}^{(2+1)}$. We first calculate the
Gaussian
part ${\cal S}_{\ell_1\ell_2}^{(2)} r_{\ell_1} r_{\ell_2}/2$ in the
action. It consists of two contributions, see Figs.~\ref{fig131}(b)
and \ref{fig132}(b). The first term is easily read off from
${\cal S}_{\text{dir}}$ in Eq.~(\ref{13A2}),
\begin{equation}
\frac12\,\left.{\cal S}_{\ell_1\ell_2}^{(2)}\right|_{\text{dir}}
  r_{\ell_1} r_{\ell_2}
  = \frac{4\pi^2}{{\cal N}} \sum_{{\bf q},i\nu_n}
   {\tilde J}\, \Delta{\bf Q}(-{\bf q},-i\nu_n)
   \cdot\Delta{\bf Q}({\bf q},i\nu_n) a^2 .
\label{13A5a}
\end{equation}
Note that
$\Delta Q_j(-{\bf q},-i\nu_n) = \Delta Q_j^\ast({\bf q},i\nu_n)$ since
$\Delta Q_j({\bf r},\tau)$ is real. The same holds for $\Delta\lambda$,
below.

\begin{figure}[htb]
\centerline{
\epsfig{file=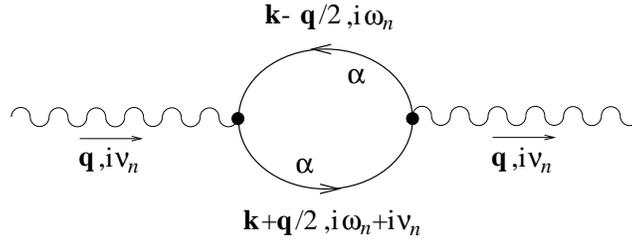,width=3.3in}}
\caption{Notation of momenta and frequencies for ${\cal S}^{(2)}$
with external $\Delta\lambda$ or $\Delta{\bf Q}$ legs.}
\label{fig133}
\end{figure}

The loop part follows from Eq.~(\ref{13A4}). The notation is shown
in Fig.~\ref{fig133}.
Inserting Eqs.~(\ref{13G1}), (\ref{13v3}), and (\ref{13v4}) we obtain
\begin{eqnarray}
\lefteqn{
\frac12\,\left.{\cal S}_{\ell_1\ell_2}^{(2)}\right|_{\text{loop}}
  r_{\ell_1} r_{\ell_2}
  = -\frac1{2N} \sum_{{\bf q},i\nu_n} \sum_{{\bf k},i\omega_n}
  \sum_\alpha } \nonumber \\
& & \quad\times \frac1{-i\beta\hbar\omega_n-i\beta\hbar\nu_n
  +{\tilde J} S ({\bf k}+{\bf q}/2)^2a^2-{\tilde B} h_\alpha^\alpha
  + {\overline{\Lambda}}}\,
  \frac{2\pi}{{\cal N}} \left[ i\beta a^2 \Delta\lambda({\bf q},i\nu_n)
    - 2{\tilde J} a^2 ({\bf k}+{\bf q}/2)
    \cdot\Delta{\bf Q}({\bf q},i\nu_n) \right] \nonumber \\
& & \quad\times \frac1{-i\beta\hbar\omega_n
    +{\tilde J} S ({\bf k}-{\bf q}/2)^2a^2-{\tilde B} h_\alpha^\alpha
    + {\overline{\Lambda}}}\,
  \frac{2\pi}{{\cal N}} \left[ i\beta a^2
  \Delta\lambda(-{\bf q},-i\nu_n)
    - 2{\tilde J} a^2 ({\bf k}-{\bf q}/2)
    \cdot\Delta{\bf Q}(-{\bf q},-i\nu_n) \right] .
\end{eqnarray}
Performing the Matsubara sum over $i\omega_n$
and utilizing the periodicity of the Bose function $n_B$ we get
\begin{eqnarray}
\frac12\,\left.{\cal S}_{\ell_1\ell_2}^{(2)}\right|_{\text{loop}}
  r_{\ell_1} r_{\ell_2}
  & = & +\frac1{2N}\,\frac{4\pi^2}{{\cal N}^2} \sum_{{\bf q},i\nu_n}
  \sum_{\bf k} \sum_\alpha
  \frac{n_B(\epsilon_{{\bf k}+{\bf q}/2}^\alpha)
  - n_B(\epsilon_{{\bf k}-{\bf q}/2}^\alpha)}
    {-i\beta\hbar\nu_n+2{\tilde J} S\,{\bf q}\cdot{\bf k} a^2}
  \nonumber \\
& & \times \left[ i\beta a^2 \Delta\lambda(-{\bf q},-i\nu_n)
  - 2{\tilde J} a^2 ({\bf k}-{\bf q}/2)
  \cdot\Delta{\bf Q}(-{\bf q},-i\nu_n) \right]  \nonumber \\
& & \times \left[ i\beta a^2 \Delta\lambda({\bf q},i\nu_n)
  - 2{\tilde J} a^2 ({\bf k}+{\bf q}/2)
  \cdot\Delta{\bf Q}({\bf q},i\nu_n) \right] ,
\end{eqnarray}
where $\epsilon_{\bf k}^\alpha \equiv {\tilde J} S\,k^2a^2
  - {\tilde B} h_\alpha^\alpha + {\overline{\Lambda}}$.
At this point we use the transverse gauge,
${\bf q}\cdot\Delta{\bf Q}({\bf q},i\nu_n)=0$. We choose coordinates
in such a way that $k_1$ and $\Delta Q_1$ are parallel to ${\bf q}$.
Then $\Delta Q_1({\bf q},i\nu_n)=0$ and the last expression simplifies,
\begin{eqnarray}
\frac12\,\left.{\cal S}_{\ell_1\ell_2}^{(2)}\right|_{\text{loop}}
  r_{\ell_1} r_{\ell_2}
  & = & \frac1{2N}\,\frac{4\pi^2}{{\cal N}^2} \sum_{{\bf q},i\nu_n}
  \sum_{\bf k} \sum_\alpha
  \frac{n_B(\epsilon_{{\bf k}+{\bf q}/2}^\alpha)
    - n_B(\epsilon_{{\bf k}-{\bf q}/2}^\alpha)}
    {-i\beta\hbar\nu_n+2{\tilde J} S\,q k_1 a^2} \nonumber \\
& & \times \Big[ -\beta^2 a^4\, \Delta\lambda(-{\bf q},-i\nu_n)
  \Delta\lambda({\bf q},i\nu_n)
  + 4{\tilde J}^2 a^4 k_2^2\, \Delta Q_2(-{\bf q},-i\nu_n)
  \Delta Q_2({\bf q},i\nu_n) \Big] .
\label{13A7}
\end{eqnarray}
Terms mixing $\Delta\lambda$ and $\Delta{\bf Q}$ vanish
since their coefficient is odd in $k_2$. Adding Eq.~(\ref{13A5a})
to Eq.~(\ref{13A7}) yields
\begin{eqnarray}
{\cal S}_{\Delta\lambda(-{\bf q},-i\nu_n),
  \Delta\lambda({\bf q},i\nu_n)}^{(2)} & = &
  \frac{4\pi^2}{{\cal N}^2 N} \sum_{\bf k} \sum_\alpha
  \frac{n_B(\epsilon_{{\bf k}+{\bf q}/2}^\alpha)
    - n_B(\epsilon_{{\bf k}-{\bf q}/2}^\alpha)}
    {-i\beta\hbar\nu_n+2{\tilde J} S\,q k_1 a^2} (-\beta^2 a^4) ,
\label{13A7a} \\
{\cal S}_{\Delta Q_2(-{\bf q},-i\nu_n),\Delta Q_2({\bf q},i\nu_n)}^{(2)}
  & = & \frac{8\pi^2}{{\cal N}}\, {\tilde J} a^2
  + \frac{4\pi^2}{{\cal N}^2 N} \sum_{\bf k} \sum_\alpha
  \frac{n_B(\epsilon_{{\bf k}+{\bf q}/2}^\alpha)
    - n_B(\epsilon_{{\bf k}-{\bf q}/2}^\alpha)}
    {-i\beta\hbar\nu_n+2{\tilde J} S\,q k_1 a^2} 4{\tilde J}^2 a^4 k_2^2
  ,
\label{13A7b}
\end{eqnarray}
all other components vanish. The fact that ${\cal S}^{(2)}$ only
connects fluctuations at $({\bf q},i\nu_n)$ and $(-{\bf q},-i\nu_n)$
just means that the RPA propagator conserves energy and momentum.
The real part of ${\cal S}^{(2)}$ is always positive except for
${\cal S}_{\Delta Q_2(0,0),\Delta Q_2(0,0)}^{(2)} = 0$. Thus there is
one zero mode,
which results in an additional factor in the partition function,
which, however, does not depend on field and is thus irrelevant for
the magnetization. The zero mode at ${\bf q}=0$, $i\nu_n=0$
shows that gauge ($\Delta{\bf Q}$)
fluctuations are massless. For the remaining modes,
${\cal S}^{(2)}$ can be inverted to get the RPA propagator $D$, which
is also positive. The saddle-point is thus stable.

Looking at the diagrams in Fig.~\ref{fig132.5} we see that the
horizontal propagator in the right diagram can only be at ${\bf q}=0$,
$i\nu_n=0$ since the source $j_\alpha$ does not insert any frequency or
momentum. In fact it can only be $\Delta\lambda(0,0)$, as we will see.
Keeping this is mind we calculate ${\cal S}_{\ell_1\ell_2\ell_3}^{(3)}$.
Since ${\cal S}^{(3)}$
is symmetric in its indices we can assume that $r_{\ell_1}$ is
$\Delta\lambda(0,0)$. Furthermore, $r_{\ell_2}$ determines $r_{\ell_3}$.
We start from the definition (\ref{13A4}),
\begin{equation}
{\cal S}_{\ell_1\ell_2\ell_3}^{(3)}
  = \frac1{N} \big[ \text{Tr} \left(G_0 \upsilon_{\ell_1}
  G_0 \upsilon_{\ell_2} G_0 \upsilon_{\ell_3}\right)
  + \text{Tr} \left(G_0 \upsilon_{\ell_1} G_0 \upsilon_{\ell_3}
  G_0 \upsilon_{\ell_2}\right) \big] .
\label{13A6}
\end{equation}
The first of the two summands is
\begin{eqnarray}
\lefteqn{
\frac1{N} \sum_{{\bf k},i\omega_n} \sum_\alpha \beta^3 \upsilon_{\ell_1}
  \upsilon_{\ell_2} \upsilon_{\ell_3}
  \frac1{-i\beta\hbar\omega_n+{\tilde J}S\,
    ({\bf k}-{\bf q}/2)^2 a^2-{\tilde B} h_\alpha^\alpha
    + {\overline{\Lambda}}} } \nonumber \\
& & \qquad\times \frac1{-i\beta\hbar\omega_n-i\beta\hbar\nu_n
    +{\tilde J} S\,({\bf k}+{\bf q}/2)^2a^2-{\tilde B} h_\alpha^\alpha
    + {\overline{\Lambda}}}\,
  \frac1{-i\beta\hbar\omega_n+{\tilde J} S\,
    ({\bf k}-{\bf q}/2)^2 a^2-{\tilde B} h_\alpha^\alpha
    + {\overline{\Lambda}}}  \nonumber \\
& & \quad = \frac1{N} \sum_{\bf k} \sum_\alpha \beta^3
  \upsilon_{\ell_1} \upsilon_{\ell_2} \upsilon_{\ell_3}
  \Bigg[ \frac{n_B({\tilde J} S\,({\bf k}+{\bf q}/2)^2a^2
    -{\tilde B} h_\alpha^\alpha + {\overline{\Lambda}})}
    {(i\beta\hbar\nu_n - 2{\tilde J} S\,q k_1 a^2)^2} \nonumber \\
& & \qquad{}- \frac{d}{dz}\,\left.\frac{n_B(z)}
  {-z-i\beta\hbar\nu_n+{\tilde J} S\,({\bf k}+{\bf q}/2)^2a^2
  - {\tilde B} h_\alpha^\alpha + {\overline{\Lambda}}}
  \right|_{z = {\tilde J} S ({\bf k}-{\bf q}/2)^2 a^2
    - {\tilde B} h_\alpha^\alpha + {\overline{\Lambda}}} \Bigg]
  \nonumber \\
& & \quad = \frac1{N} \sum_{\bf k} \sum_\alpha \beta^3 \upsilon_{\ell_1}
  \upsilon_{\ell_2} \upsilon_{\ell_3} \Bigg[ \frac{n_B({\tilde J} S\,
    ({\bf k}+{\bf q}/2)^2a^2-{\tilde B} h_\alpha^\alpha
    + {\overline{\Lambda}})}
    {(-i\beta\hbar\nu_n + 2{\tilde J} S\,q k_1 a^2)^2} \nonumber \\
& & \qquad{}- \frac{n_B({\tilde J} S\,({\bf k}-{\bf q}/2)^2 a^2
  -{\tilde B} h_\alpha^\alpha + {\overline{\Lambda}})}
    {(-i\beta\hbar\nu_n + 2{\tilde J} S\,q k_1 a^2)^2}
  - \frac{n_B^{(1)}({\tilde J} S\,({\bf k}-{\bf q}/2)^2 a^2
  -{\tilde B} h_\alpha^\alpha + {\overline{\Lambda}})}
    {-i\beta\hbar\nu_n + 2{\tilde J} S\,q k_1 a^2} \Bigg] ,
\end{eqnarray}
where $n_B^{(\nu)}(\epsilon) \equiv d^\nu n_B(\epsilon)/d\epsilon^\nu$
is the $\nu$-th derivative of the Bose function. With the vertex factors
the last expression becomes
\begin{eqnarray}
\ldots & = & \frac{(2\pi)^3}{{\cal N}^3 N} \sum_{\bf k}
  \sum_\alpha i\beta a^2\,
  \left\{{-\beta^2 a^4 \atop 4{\tilde J}^2 a^4 k_2^2} \right\}
  \Bigg[ \frac{n_B({\tilde J} S ({\bf k}+{\bf q}/2)^2a^2
    -{\tilde B} h_\alpha^\alpha + {\overline{\Lambda}})}
    {(-i\beta\hbar\nu_n + 2{\tilde J} S\, q k_1 a^2)^2} \nonumber \\
& & {}- \frac{n_B({\tilde J} S ({\bf k}-{\bf q}/2)^2a^2
    -{\tilde B} h_\alpha^\alpha + {\overline{\Lambda}})}
    {(-i\beta\hbar\nu_n + 2{\tilde J} S\, q k_1 a^2)^2}
  - \frac{n_B^{(1)}({\tilde J} S ({\bf k}-{\bf q}/2)^2a^2
    -{\tilde B} h_\alpha^\alpha + {\overline{\Lambda}})}
    {-i\beta\hbar\nu_n + 2{\tilde J} S\, q k_1 a^2} \Bigg] ,
\label{13A6a}
\end{eqnarray}
where the upper [lower] term in the curly brackets is for
$r_{\ell_2} = \Delta\lambda({\bf q},i\nu_n)$
[$\Delta Q_2({\bf q},i\nu_n)$]. For $r_{\ell_1} = \Delta Q_2(0,0)$
the integrand would be odd in $k_2$ so that this contribution vanishes.

The second term in Eq.~(\ref{13A6}) just has $r_{\ell_2}$ and
$r_{\ell_3}$ exchanged,
which means ${\bf q}$ and $i\nu_n$ have opposite sign. Thus,
\begin{eqnarray}
\!\!{\cal S}_{\Delta\lambda(0,0),\Delta\lambda(-{\bf q},-i\nu_n),
  \Delta\lambda({\bf q},i\nu_n)}^{(3)}\!
  & = & \frac{(2\pi)^3}{{\cal N}^3 N}\,i\beta a^2 \sum_{\bf k}
  \sum_\alpha \frac{n_B^{(1)}(\epsilon_{{\bf k}+{\bf q}/2}^\alpha)
    - n_B^{(1)}(\epsilon_{{\bf k}-{\bf q}/2}^\alpha)}
    {-i\beta\hbar\nu_n+2{\tilde J} S\,q k_1 a^2} (-\beta^2 a^4) ,
\label{13A7c} \\
\!\!{\cal S}_{\Delta\lambda(0,0),\Delta Q_2(-{\bf q},-i\nu_n),
  \Delta Q_2({\bf q},i\nu_n)}^{(3)}\!
  & = & \frac{(2\pi)^3}{{\cal N}^3 N}\,i\beta a^2 \sum_{\bf k}
  \sum_\alpha \frac{n_B^{(1)}(\epsilon_{{\bf k}+{\bf q}/2}^\alpha)
    - n_B^{(1)}(\epsilon_{{\bf k}-{\bf q}/2}^\alpha)}
    {-i\beta\hbar\nu_n+2{\tilde J} S\,q k_1 a^2} 4{\tilde J}^2 a^4 k_2^2
  ,
\label{13A7d}
\end{eqnarray}
where the fractions in Eq.~(\ref{13A6a}) containing the denominator
squared have cancelled upon adding the two terms.

We can now calculate ${\cal S}^{(1+1)}$ and ${\cal S}^{(2+1)}$. These
expressions
contain a vertex $\upsilon_{j_\alpha} = 2\pi a^2/{\cal N}$ instead
of $\upsilon_{\Delta\lambda} = 2\pi i a^2/{\cal N}$.
The source $j_\alpha$ inserts zero momentum and frequency.
For ${\cal S}^{(1+1)}$ we are thus only interested in
${\cal S}_{j_\alpha;\Delta\lambda(0,0)}^{(1+1)}$ [the left loop in
Fig.~\ref{fig132.5}(b)]. By taking the limit to zero frequency and
momentum, we obtain
\begin{eqnarray}
{\cal S}_{\Delta\lambda(0,0),\Delta\lambda(0,0)}^{(2)} & = &
  \frac{4\pi^2}{{\cal N}^2 N} \sum_{\bf k} \sum_\alpha 
  n_B^{(1)}({\tilde J} S k^2 a^2 - {\tilde B} h_\alpha^\alpha
    + {\overline{\Lambda}}) (-\beta^2 a^4)  \nonumber \\
& = & \frac{\pi}{2{\cal N} {\tilde J} S}\,\beta^2 a^4 \left[
  n_B({\overline{\Lambda}}-{\tilde B})
  + n_B({\overline{\Lambda}}+{\tilde B})\right] .
\label{13A7aa}
\end{eqnarray}
Keeping in mind that $j_\alpha$ couples only to the boson of flavor
$\alpha$ we find similarly
\begin{equation}
{\cal S}_{j_\alpha;\Delta\lambda(0,0)}^{(1+1)}
  = -\frac{i\pi}{{\cal N} N {\tilde J} S}\,\beta^2 a^4\,
  n_B({\overline{\Lambda}}-{\tilde B} h_\alpha^\alpha) .
\label{13A7a1}
\end{equation}
From ${\cal S}^{(3)}$ we infer ${\cal S}^{(2+1)}$,
\begin{eqnarray}
{\cal S}_{j_\alpha;\Delta\lambda(-{\bf q},-i\nu_n),
  \Delta\lambda({\bf q},i\nu_n)}^{(2+1)}
  & = & \frac{(2\pi)^3}{{\cal N}^3 N}\,\beta a^2 \sum_{\bf k}
  \frac{n_B^{(1)}(\epsilon_{{\bf k}+{\bf q}/2}^\alpha)
    - n_B^{(1)}(\epsilon_{{\bf k}-{\bf q}/2}^\alpha)}
    {-i\beta\hbar\nu_n+2{\tilde J} S\,q k_1 a^2} (-\beta^2 a^4) ,
\label{13A7c1} \\
{\cal S}_{j_\alpha;\Delta Q_2(-{\bf q},-i\nu_n),\Delta Q_2({\bf q},
  i\nu_n)}^{(2+1)}
& = & \frac{(2\pi)^3}{{\cal N}^3 N}\,\beta a^2 \sum_{\bf k}
  \frac{n_B^{(1)}(\epsilon_{{\bf k}+{\bf q}/2}^\alpha)
    - n_B^{(1)}(\epsilon_{{\bf k}-{\bf q}/2}^\alpha)}
    {-i\beta\hbar\nu_n+2{\tilde J} S\,q k_1 a^2} 4{\tilde J}^2 a^4 k_2^2
  .
\label{13A7d1}
\end{eqnarray}

\section{Calculation of diagrams for O($N$)}
\label{appB}

\begin{figure}[htb]
\centerline{
\epsfig{file=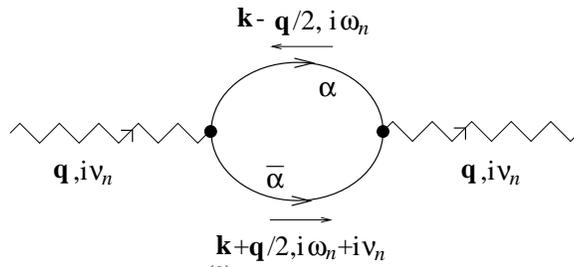,width=3in}}
\caption{Notation of momenta and frequencies for ${\cal S}^{(2)}$ with
external $\Delta\mu$ legs. The momenta and frequencies are measured in
the direction of the attached arrows.}
\label{fig231}
\end{figure}

We start with ${\cal S}^{(2)}$, using the notation shown in
Fig.~\ref{fig231}. In analogy to Eq.~(\ref{13A4}),
\begin{equation}
{\cal S}_{\Delta\mu^\ast({\bf q},i\nu_n),
  \Delta\mu({\bf q},i\nu_n)}^{(2)}
  = -\frac1{N}\,\text{Tr} \left( G_0 \upsilon_{\Delta\mu^\ast}
  G_0^\ast \upsilon_{\Delta\mu} \right) .
\end{equation}
Here, one of the Greens functions is the complex conjugate since the
line is traversed against the direction of the boson propagator.
Momentum and frequency of $G_0^\ast$ are measured counter-clockwise.
We find
\begin{eqnarray}
{\cal S}_{\Delta\mu^\ast({\bf q},i\nu_n),
  \Delta\mu({\bf q},i\nu_n)}^{(2)}
& = & -\frac{\pi^2}{{\cal N}^2 N}\,\beta^2a^4 \sum_{{\bf k},i\omega_n}
  \sum_\alpha \frac1{-i\beta\hbar\omega_n-i\beta\hbar\nu_n
    + {\tilde J} S ({\bf k}+{\bf q}/2)^2a^2
    - {\tilde B}\hat{h}_\alpha^\alpha+{\overline{\Lambda}}} \nonumber \\
& & \times \frac1{-i\beta\hbar\omega_n-{\tilde J} S
  ({\bf k}-{\bf q}/2)^2a^2
  + {\tilde B}\hat{h}_{\overline{\alpha}}^{\overline{\alpha}}
  - {\overline{\Lambda}}}  \nonumber \\
& = & \frac{\pi^2}{{\cal N}^2 N}\,\beta^2a^4 \sum_{\bf k} \sum_\alpha
  \frac{1+n_B(\epsilon_{{\bf k}+{\bf q}/2}^\alpha)
    + n_B(\epsilon_{{\bf k}-{\bf q}/2}^{\overline{\alpha}})}
    {-i\beta\hbar\nu_n + 2{\tilde J} S\,k^2a^2
    + {\tilde J} S\,q^2a^2/2 + 2{\overline{\Lambda}}}
\end{eqnarray}
with $\epsilon_{\bf k}^\alpha
  \equiv {\tilde J} S\,k^2a^2 - {\tilde B}\hat{h}_\alpha^\alpha
  + {\overline{\Lambda}}$. Here,
we have used that
$\hat{h}_\alpha^\alpha+\hat{h}_{\overline{\alpha}}^{\overline{\alpha}}
  =0$
and the identity $n_B(-\epsilon) = -n_B(\epsilon)-1$. The real part 
is positive so that the functional integral is well-defined. This kind
of expression is known from the theory of scattering processes.

${\cal S}_{\Delta\lambda(0,0),\Delta\mu^\ast,\Delta\mu}^{(3)}$ can
be derived similarly,
\begin{eqnarray}
{\cal S}_{\Delta\lambda(0,0),\Delta\mu^\ast({\bf q},i\nu_n),
  \Delta\mu({\bf q},i\nu_n)}^{(3)}
  & = & \frac{(2\pi)^3}{{\cal N}^3 N}\,\frac{i\beta^3 a^6}{4}
  \sum_{\bf k} \sum_\alpha
  \Bigg( \frac{n_B^{(1)}(\epsilon_{{\bf k}+{\bf q}/2}^\alpha) +
    n_B^{(1)}(\epsilon_{{\bf k}-{\bf q}/2}^{\overline{\alpha}})}
  {-i\beta\hbar\nu_n + 2{\tilde J} S\,k^2a^2 + {\tilde J} S\,q^2a^2/2
  + 2{\overline{\Lambda}}} \nonumber \\
& & {}- 2\frac{1+n_B(\epsilon_{{\bf k}+{\bf q}/2}^\alpha)
  + n_B(\epsilon_{{\bf k}-{\bf q}/2}^{\overline{\alpha}})}
    {(-i\beta\hbar\nu_n + 2{\tilde J} S\,k^2a^2
    + {\tilde J} S\,q^2a^2/2 + 2{\overline{\Lambda}})^2} \Bigg) .
\label{23A9}
\end{eqnarray}
The term containing the denominator squared does not cancel in this
case. To obtain this result we have summed over boson frequencies
$i\omega_n$ using contour integration. One has to consider operator
ordering to do this properly. Since the anomalous combinations
$d_{\overline{\alpha}}^\dagger d_\alpha^\dagger$ and
$d_{\overline{\alpha}} d_\alpha$ in the Hamiltonian contain two
commuting operators time splitting is not necessary and no phase
factors appear in $\upsilon_{\Delta\mu}$ and
$\upsilon_{\Delta\mu^\ast}$. On the other hand,
the other vertices obtain factors $\exp(i\omega_n\eta)$.
It can be shown that the two terms in ${\cal S}^{(3)}$, coming from the
symmetrization in Eq.~(\ref{13A6}), obtain factors of
$\exp(i\omega_n\eta)\exp(i\nu_n\eta) $ and
$\exp(-i\omega_n\eta)\exp(i\nu_n\eta)$, respectively. The different
factors in
$i\omega_n$ are crucial in arriving at Eq.~(\ref{23A9}). Furthermore,
we obtain an overall factor of $\exp(i\nu_n\eta)$.

Immediately we find
\begin{eqnarray}
{\cal S}_{j_\alpha;\Delta\mu^\ast({\bf q},i\nu_n),\Delta\mu({\bf q},
  i\nu_n)}^{(2+1)}
  & = & \frac{(2\pi)^3}{{\cal N}^3 N}\,\frac{\beta^3 a^6}{4}
  \sum_{\bf k} \Bigg(
  \frac{n_B^{(1)}(\epsilon_{{\bf k}+{\bf q}/2}^\alpha)
    + n_B^{(1)}(\epsilon_{{\bf k}-{\bf q}/2}^\alpha)}
    {-i\beta\hbar\nu_n + 2{\tilde J} S\,k^2a^2
    + {\tilde J} S\,q^2a^2/2 + 2{\overline{\Lambda}}} \nonumber \\
& & {}- 2\frac{1+n_B(\epsilon_{{\bf k}+{\bf q}/2}^\alpha)
    + n_B(\epsilon_{{\bf k}-{\bf q}/2}^\alpha)}
    {(-i\beta\hbar\nu_n + 2{\tilde J} S\,k^2a^2
    + {\tilde J} S\,q^2a^2/2 + 2{\overline{\Lambda}})^2} \Bigg) .
\end{eqnarray}
Finally, we have to recalculate
${\cal S}_{\Delta\lambda(0,0),\Delta\lambda(0,0)}^{(2)}$. By replacing
$h_\alpha^\alpha$ by $\hat{h}_\alpha^\alpha$ in Eq.~(\ref{13A7aa})
we get
\begin{equation}
{\cal S}_{\Delta\lambda(0,0),\Delta\lambda(0,0)}^{(2)}
  = \frac{\pi}{3{\cal N} {\tilde J} S}\,\beta^2 a^4
  \left[n_B({\overline{\Lambda}}-{\tilde B})
  + n_B({\overline{\Lambda}})+n_B({\overline{\Lambda}}+{\tilde B})
  \right] .
\end{equation}




\end{document}